# Bayesian FFT Modal Identification for Multi-setup Experimental Modal Analysis

Peixiang Wang [a,b], Binbin Li [a,b,*]

[a] ZJU-UIUC Institute, Zhejiang University, Haining, Zhejiang, China

[b] College of Civil Engineering and Architecture, Zhejiang University, Hangzhou, Zhejiang, China

## Abstract

In full-scale forced vibration tests, the demand often arises to capture high-spatial-resolution mode shapes with limited number of sensors and shakers. Multi-setup experimental modal analysis (EMA) addresses this challenge by roving sensors and shakers across multiple setups. To enable fast and accurate multi-setup EMA, this paper develops a Bayesian modal identification strategy by extending an existing single-setup algorithm. Specifically, a frequency-domain probabilistic model is first formulated using multiple sets of structural multiple-input, multiple-output (MIMO) vibration data. A constrained Laplace method is then employed for Bayesian posterior approximation, providing the maximum a posteriori estimates of modal parameters along with a posterior covariance matrix (PCM) for uncertainty quantification. Utilizing complex matrix calculus, analytical expressions are derived for parameter updates in the coordinate descent optimization, as well as for PCM computation, enhancing both coding simplicity and computational efficiency. The proposed algorithm is intensively validated by investigating empirical examples with synthetic and field data. It demonstrates that the proposed method yields highly consistent results compared to scenarios with adequate test equipment. The resulting high-fidelity MIMO model enables structural response prediction under future loading conditions and supports condition assessment.

# 1 Introduction

Field vibration test provides a practical way to acquire in-situ data from a constructed structure. The recorded dynamic input and responses can be used to extract structural modal properties (e.g., natural frequencies, damping ratios, mode shapes), which play a critical role in the field of structural health monitoring [1,2], model updating [3,4] and vibration control [5,6]. Owing to the advances in sensing and excitation techniques, an increasing number of free-, forced- and ambient vibration tests have been successfully conducted on various kinds of structures [7–10]. The choice of vibration test depends on the required precision of the interested modal parameters, the desired mode shapes, as well as economic factors limited by the quality and quantity of shakers/sensors.

Depending on whether the input is measured or not, vibration tests can be categorized into experimental modal analysis (EMA) [11] and operational modal analysis (OMA) [12]. Since the dynamic load and structural responses are both used in EMA, a full input-output dynamic model can be constructed, facilitating the downstream applications, e.g., the seismic response prediction [13]. For modal identification, EMA generally yields more precise results, as evidenced by empirical studies [14,15], and a recent analytical analysis, named as 'uncertainty laws' [16]. The uncertainty laws reveal that the identification uncertainties of natural frequencies and damping ratios are non-vanishing unless the input force is known, but knowing the input does not improve the identification precision of mode shapes.

In full-scale forced vibration test, obtaining high resolution mode shapes is a key task, because they carry the most spatial information, but keeps challenging [17–19]. Firstly, the number of sensors is limited, due to limited instrumentation budget and manpower. Secondly, not all modes of interest can be well excited in a single-setup test because of, e.g., the improper configuration of shakers. Thirdly, the prevalent use of a single-source shaker yields an single-input and multiple-output (SIMO) scenario, failing to capture the coupling behavior of mode shapes, e.g., in symmetrical buildings [20] and long cable-stayed bridges [21]. A more feasible strategy is to conduct a multi-setup vibration test, typically with each setup covering different locations of the structure but sharing some reference DOFs. It should be noted that only quantity or location of sensors changes across different setups in the multi-setup OMA, while it involves the changing of both sensors and shakers in multi-setup EMA.

The mode shape in the multi-setup setting is often called the 'global' mode shape, because it is assembled from various 'local' mode shapes measured in each setup. Conventionally, the global mode shape can be obtained through either pre- or post-identification approach. Pre-identification approach refers to merging datasets from different setups before modal identification [22–24]. Often, it imposes a priori assumptions that all modal parameters remain time-invariant across all setups, which could potentially induce modeling error, particularly in ambient vibration tests. Post-identification approach allows initially identifying modal parameters using the individual data from each setup, and assemble the local mode shapes to the targeted global one [25,26].

Various methods have been proposed for global mode shape assembly in multi-setup OMA, e.g., local and global least squares methods [12]. Although they perform well if the local mode shapes in all setups are well-identified, a study [27] shows that the Bayesian method tends to be more robust in cases where significant disagreement exists across different setups. Another notable benefit of Bayesian approach lies in its capability to quantify the estimation uncertainty, which may be critical for downstream applications. Owing to these potential benefits, we choose to work on a Bayesian approach for multi-setup EMA. In this aspect, Au and his colleagues developed a series of fast algorithms [28,29] and analytical uncertainty analysis [30] in addressing multi-setup OMA problem, casted into the framework of BAYOMA (Bayesian operational modal analysis). For the single-setup EMA problem, Ni et al. developed a series of algorithm for the case of single-source shaker excitation [31–33] and the case of single-source seismic excitation [34,35]; Wang et al. developed a Bayesian fast Fourier transform (FFT) algorithm for the general multiple-input, multiple-output (MIMO) EMA [13]. However, the multi-setup EMA has not got enough attention, because it is regarded to be impractical for civil engineering structures. In this paper, we will show that a small shaker with a moving mass of 30 kg is able to excite a building/bridge with a modal mass of 10 ton to an acceptable magnitude for identification, illustrating the feasibility of the multi-setup EMA for small to moderate size of buildings/bridges. Since the multi-setup EMA can build a high-fidelity input-output model, e.g., used for damage diagnosis and prognosis, it may play an important role in the maintenance of small to moderate size of buildings/bridges.

In this paper, we present a fast Bayesian FFT algorithm for multi-setup EMA of civil engineering structures, considering the changing configuration of sensors and shakers. The remaining paper is organized as follows. A probabilistic model of multi-setup EMA is formulated in Section 2, leveraging a linear, time-invariant and classically damped physical model. In Section 3, a constrained Laplace approximation is introduced to efficiently compute the posterior statistics, including the maximum a posteriori (MAP) estimate and the corresponding posterior covariance matrix (PCM). Finally, empirical studies are presented in Section 4 with synthetic and field data to fully illustrate the performance of the developed algorithm.

## 2   Probabilistic model of multi-setup EMA

This section presents the probabilistic model for modal identification using multiple sets of MIMO vibration data. This includes outlining the modeling assumptions, specifying the parameters to be identified, and formulating the likelihood function.

In a multi-setup EMA consisting of $n_r$ different configuration of sensors, let $\left\{\hat{\boldsymbol{y}}_j^{(r)} \in \mathbb{R}^{d_r}\right\}_{j=0}^{N_r-1}$ denote the measured time history in Setup $r$ ($r = 1, \ldots, n_r$) where $N_r$ and $d_r$ are the number of samples per data channel and the number of data channels, respectively. The (two-sided) scaled FFT of $\hat{\boldsymbol{y}}_j^{(r)}$ at frequency $f_k^{(r)} = k/(N_r \Delta t_r)$ (Hz) is defined as

$$\hat{\boldsymbol{Y}}_k^{(r)} = \sqrt{\frac{\Delta t_r}{N_r}} \sum_{j=0}^{N_r-1} \hat{\boldsymbol{y}}_j^{(r)} \exp\left(-\frac{\mathbf{i} 2\pi jk}{N_r}\right) \tag{1}$$

where $\mathbf{i}^2 = -1$ and $\Delta t_r$ (sec) is the sampling interval. The $\hat{\boldsymbol{Y}}_k^{(r)}$ has been appropriately scaled such that the $\mathrm{E}[\hat{\boldsymbol{Y}}_k^{(r)} \hat{\boldsymbol{Y}}_k^{*(r)}]$ ($\mathrm{E}[\cdot]$ denotes "expectation") yields the two-sided sample power spectral density (PSD). Denoting $\boldsymbol{Y}_k^{(r)}$ as the scaled FFT (as defined in Eqn. (1)) of the theoretical structural dynamic, we assume

$$\hat{\boldsymbol{Y}}_k^{(r)} = \boldsymbol{Y}_k^{(r)} + \boldsymbol{\varepsilon}_k^{(r)} \tag{2}$$

where $\boldsymbol{\varepsilon}_k^{(r)}$ denotes the scaled FFT of an error term that accounts for measurement noise and modeling error at frequency $f_k^{(r)}$ in Setup $r$. It is assumed that $\boldsymbol{\varepsilon}_k^{(r)}$ follows a complex Gaussian distribution with zero mean and covariance matrix of $S_e^{(r)} \boldsymbol{I}_{d_r}$ ($\boldsymbol{I}_{d_r} \in \mathbb{R}^{d_r \times d_r}$ denotes the identity matrix) within the selected band, i.e., $\boldsymbol{\varepsilon}_k^{(r)} \sim \mathcal{CN}(0, S_e^{(r)} \boldsymbol{I}_{d_r})$, where $S_e^{(r)}$ is also called the theoretical PSD of the error $\boldsymbol{\varepsilon}_k^{(r)}$.

For a discretized, linear, time-invariant dynamical system with the number of degrees of freedom (DOFs) being $n$, the structural dynamic responses under a shaker excitation can be calculated by solving the equation of motion

$$\boldsymbol{M}\ddot{\boldsymbol{y}}(t) + \boldsymbol{C}\dot{\boldsymbol{y}}(t) + \boldsymbol{K}\boldsymbol{y}(t) = \boldsymbol{L}_0^{(s)} m_s \ddot{\boldsymbol{u}}^{(r)}(t) \tag{3}$$

where $\boldsymbol{M} \in \mathbb{R}^{n \times n}$, $\boldsymbol{C} \in \mathbb{R}^{n \times n}$, and $\boldsymbol{K} \in \mathbb{R}^{n \times n}$ are the structural mass, damping and stiffness matrices, respectively; $\ddot{\boldsymbol{y}}(t) \in \mathbb{R}^n$, $\dot{\boldsymbol{y}}(t) \in \mathbb{R}^n$, and $\boldsymbol{y}(t) \in \mathbb{R}^n$ are the nodal acceleration, velocity and displacement responses, respectively; the shaker force is defined as the production of the nominal mass $m_s$ and the input acceleration $\ddot{\boldsymbol{u}}^{(r)}(t) \in \mathbb{R}^{d_s}$ ($d_s$ is the number of external excitations); the superscript $s$ indicates the $s$-th setup of shaker configuration, where the shaker location is determined by the selection matrix $\boldsymbol{L}_0^{(s)} \in \mathbb{R}^{n \times d_s}$. The $(j,k)$-entry of $\boldsymbol{L}_0^{(s)}$ is 1 if the $k$-th component of excitations is applied at DOF $j$ of the structure and zero otherwise. Without loss of generality, we assume the total number of shaker placement schemes $n_s$ is less than that of sensor placement schemes $n_r$.

Assuming classical damping and taking advantage of the narrow band behavior of structural vibration, one can approximate the frequency-domain structural dynamic responses within a selected frequency band $[f_l, f_u]$ as [12]

$$\boldsymbol{Y}_k^{(r)} \approx \sum_{i=1}^{m} \boldsymbol{S}_0^{(r)} \boldsymbol{\psi}_i \eta_{ik}^{(r)} \tag{4}$$

following the principle of modal superposition. The mode shape vector $\boldsymbol{\psi}_i \in \mathbb{R}^n$ of the $i$-th mode is the generalized eigenvector of the mass matrix $\boldsymbol{M}$ and the stiffness matrix $\boldsymbol{K}$. The selection matrix $\boldsymbol{S}_0^{(r)} \in \mathbb{R}^{d_r \times n}$ represents sensor configuration in the $r$-th setup. The $(j,k)$-entry of $\boldsymbol{S}_0^{(r)}$ is 1 if the $j$-th data channel measures DOF $k$ and zero otherwise. Here, the number of dominated modes $m$ in the selected band is typically no more than 3 in practical civil engineering structures. This band-by-band modeling strategy directly improves computational efficiency by ignoring less informative data. The modal response $\eta_{ik}^{(r)}$ can be expressed as

$$\eta_{ik}^{(r)} = h_{ik}^{(r)} \boldsymbol{\lambda}_i^{(s)\mathrm{T}} \ddot{\boldsymbol{U}}_k^{(r)} \tag{5}$$

where $\ddot{\boldsymbol{U}}_k^{(r)}$ is the scaled FFT of $\ddot{\boldsymbol{u}}^{(r)}(t)$, and the frequency response function (FRF) $h_{ik}^{(r)}$ is defined as

$$h_{ik}^{(r)} = \frac{\left(2\pi \mathrm{i} f_k^{(r)}\right)^{-q}}{\left(1 - \beta_{ik}^{(r)2}\right) - \mathrm{i}\left(2\zeta_i^{(r)} \beta_{ik}^{(r)}\right)}; \quad \beta_{ik}^{(r)} = \frac{f_i}{f_k^{(r)}}; \quad q = \begin{cases} 0, & \text{acceleration data} \\ 1, & \text{velocity data} \\ 2, & \text{displacement data} \end{cases} \tag{6}$$

Here, $f_i$ and $\zeta_i$ respectively represents the natural frequency (in Hz) and damping ratio of the $i$-th mode. The symbol $\boldsymbol{\lambda}_i^{(s)\mathrm{T}} = \mu_i \boldsymbol{\psi}_i^\mathrm{T} \boldsymbol{L}_0^{(s)}$ is the $i$-th modal participation factor (MPF) in the corresponding setups, where $\mu_i = m_s / \boldsymbol{\psi}_i^\mathrm{T} \boldsymbol{M} \boldsymbol{\psi}_i$ is defined as the $i$-th modal mass ratio (MMR).

The mode shape $\boldsymbol{\psi}_i$ can be categorized into three types: $\boldsymbol{\varphi}_i \in \mathbb{R}^d$ corresponding to all DOFs instrumented with sensors, $\boldsymbol{\varphi}_i' \in \mathbb{R}^{d_1}$ corresponding to all DOFs excited by shakers, and $\boldsymbol{\varphi}_i'' \in \mathbb{R}^{d_2}$ corresponding to the remaining DOFs. With this setting, we have the equivalent expressions $\boldsymbol{S}_0^{(r)} \boldsymbol{\psi}_i = \boldsymbol{S}^{(r)} \boldsymbol{\varphi}_i$ and $\boldsymbol{\psi}_i^\mathrm{T} \boldsymbol{L}_0^{(s)} = \boldsymbol{\varphi}_i'^\mathrm{T} \boldsymbol{L}^{(s)}$, where $\boldsymbol{S}^{(r)} \in \mathbb{R}^{d_r \times d}$ retains all columns of $\boldsymbol{S}_0^{(r)}$ corresponding to the measured DOFs and $\boldsymbol{L}^{(s)} \in \mathbb{R}^{d_1 \times d_s}$ retains all rows of $\boldsymbol{L}_0^{(s)}$ corresponding to the excited DOFs. We further assume that there is no overlap between $\boldsymbol{\varphi}_i$ and $\boldsymbol{\varphi}_i'$ to reduce the estimation difficulty, although it induces loss of information if a DOF is both measured and excited. Since $\boldsymbol{\varphi}_i''$ is not related to the measured input or output, it becomes unidentifiable in the multi-setup EMA. A direct consequence is that $\boldsymbol{\varphi}_i'$ is not identifiable, because it always appears with the MMR $\mu_i$, and one can only identify their product, i.e., $\boldsymbol{\lambda}_i^{(s)}$. Rewriting the above in a matrix form, one has

$$\boldsymbol{Y}_k^{(r)} = \boldsymbol{S}^{(r)} \boldsymbol{\Phi} \boldsymbol{H}_k^{(r)} \boldsymbol{\Lambda}^{(s)\mathrm{T}} \ddot{\boldsymbol{U}}_k^{(r)} \tag{7}$$

where $\boldsymbol{\Phi} = [\boldsymbol{\varphi}_1, \boldsymbol{\varphi}_2, \ldots, \boldsymbol{\varphi}_m] \in \mathbb{R}^{d \times m}$ is the mode shape matrix confined to the measured DOFs; $\boldsymbol{H}_k^{(r)} = \mathrm{diag}[h_{1k}^{(r)}, h_{2k}^{(r)}, \ldots, h_{mk}^{(r)}] \in \mathbb{R}^{m \times m}$ is a diagonal matrix with FRFs as its diagonal elements; $\boldsymbol{\Lambda}^{(s)} = [\boldsymbol{\lambda}_1^{(s)}, \boldsymbol{\lambda}_2^{(s)}, \ldots, \boldsymbol{\lambda}_m^{(s)}] \in \mathbb{R}^{d_s \times m}$ is the MPF matrix.

For simplicity, we collect all unknown modal parameters in one frequency band into a random vector $\boldsymbol{\theta} = \left\{ \{f_i, \zeta_i, \{\boldsymbol{\lambda}_i^{(s)}\}_{s=1}^{n_s}\}_{i=1}^{m}, \{S_e^{(r)}\}_{r=1}^{n_r}, \boldsymbol{\Phi} \right\}$. Here, we assume that modal parameters are time-invariant during the test period. This is different from the assumption made in the Bayesian multi-setup OMA [28]. Comparing to multi-setup OMA, the test period of multi-setup EMA can be much shorter for a targeted identification precision [16], so that the time-variance of modal frequencies (e.g., due to changing temperature) can be ignored. In addition, since the external excitation is under control in EMA, the modal force PSD can be kept relatively constant, and thus one can neglect the amplitude-dependence of damping ratios [36]. Finally, if the shaker is placed improperly (e.g., close to the node of the mode shape), the time-variant setting may leave some modal parameters unidentifiable.

Assuming statistical independence of $\boldsymbol{\varepsilon}_k^{(r)}$ at different frequencies, it finally yields an independent complex Gaussian distribution for the measurement $\{\widehat{\boldsymbol{Y}}_k^{(r)}\}$ with the mean of $\boldsymbol{S}^{(r)}\boldsymbol{\Phi}\boldsymbol{H}_k^{(r)}\boldsymbol{\Lambda}^{(s)\mathrm{T}}\ddot{\boldsymbol{U}}_k^{(r)}$ and the covariance matrix of $S_e^{(r)}\boldsymbol{I}_{d_r}$, i.e., $\widehat{\boldsymbol{Y}}_k^{(r)} \sim \mathcal{CN}(\boldsymbol{S}^{(r)}\boldsymbol{\Phi}\boldsymbol{H}_k^{(r)}\boldsymbol{\Lambda}^{(s)\mathrm{T}}\ddot{\boldsymbol{U}}_k^{(r)}, S_e^{(r)}\boldsymbol{I}_{d_r})$. This assumption is robust by the theoretical argument that the scaled FFTs $\{\widehat{\boldsymbol{Y}}_k^{(r)}\}$ are independent and asymptotically normally distributed for long data (i.e., $N_f \gg 1$) [12]. We further assume that excitations and errors in different setups are independent, which approximately holds because there is usually a large time gap (far more than the natural period) between successive setups. This implies that the scaled FFTs $D = \{\{\widehat{\boldsymbol{Y}}_k^{(1)}\}, \dots, \{\widehat{\boldsymbol{Y}}_k^{(n_r)}\}\}$ within the selected band in all setups are all statistically independent, yielding $p(D|\boldsymbol{\theta}) = \prod_{r=1}^{n_r} p(\{\widehat{\boldsymbol{Y}}_k^{(r)}\}|\boldsymbol{\theta}) = \exp[-\mathcal{L}(\boldsymbol{\theta})]$, where the negative log-likelihood function (NLLF) $\mathcal{L}(\boldsymbol{\theta})$ is expressed as

$$\mathcal{L}(\boldsymbol{\theta}) = \sum_{r=1}^{n_r} d_r N_f^{(r)} \ln \pi + \sum_{r=1}^{n_r} d_r N_f^{(r)} \ln S_e^{(r)} + \sum_{r=1}^{n_r} S_e^{(r)-1} \sum_{k=1}^{N_f^{(r)}} \left[ (\widehat{\boldsymbol{Y}}_k^{(r)} - \boldsymbol{Y}_k^{(r)})^* (\widehat{\boldsymbol{Y}}_k^{(r)} - \boldsymbol{Y}_k^{(r)}) \right] \qquad (8)$$

The problem so far is not identifiable since the modal participation factor always accompanies the mode shape. To be more specific, identical NLLFs can be obtained by scaling $\boldsymbol{\phi}_i$ with an arbitrary constant $c$ and subsequently dividing $\boldsymbol{\lambda}_i^{(s)}$ by $c$, i.e., $\boldsymbol{\varphi}_i \boldsymbol{\lambda}_i^{(s)\mathrm{T}} = (c\boldsymbol{\varphi}_i)(\boldsymbol{\lambda}_i^{(s)\mathrm{T}}/c)$. This means that there are infinite number of minima of NLLF. A conventional treatment is to introduce unit norm constraints on mode shapes as

$$G_i(\boldsymbol{\theta}) = \boldsymbol{\varphi}_i^\mathrm{T} \boldsymbol{\varphi}_i - 1 = 0 \quad \text{for } i = 1,2, \dots, m \qquad (9)$$

This constraint renders the problem globally identifiable in the physical sense, as the sign of $\boldsymbol{\varphi}_i$ can be positive or negative without physical distinction.

## 3  Bayesian inference of modal parameters

The posterior distribution of the unknown parameter $\boldsymbol{\theta}$ is investigated in the section following the Bayes' theorem. Since the data size is generally large enough, i.e., $N_f \gg 1$, we simply assume a uniform prior distribution for $\boldsymbol{\theta}$ so that the posterior distribution is mainly determined by data, and expressed as $p(\boldsymbol{\theta}|D) \propto p(D|\boldsymbol{\theta}) = \exp[-\mathcal{L}(\boldsymbol{\theta})]$. The closed-form expression of posterior PDF $p(\boldsymbol{\theta}|D)$ cannot be obtained due to the complicated nature of NLLF $\mathcal{L}(\boldsymbol{\theta})$. With large data in multi-setup EMA, the Laplace approximation is adopted to approximate the posterior distribution [37], by fitting the posterior PDF with a local normal distribution whose mean vector $\widehat{\boldsymbol{\theta}}$ and covariance matrix $\widehat{\boldsymbol{\Sigma}}$ are respectively given by

$$\widehat{\boldsymbol{\theta}} = \arg\min_{\boldsymbol{\theta}} \mathcal{L}(\boldsymbol{\theta}); \quad \widehat{\boldsymbol{\Sigma}} = [\nabla^2 \mathcal{L}(\boldsymbol{\theta})|_{\boldsymbol{\theta}=\widehat{\boldsymbol{\theta}}}]^\dagger \qquad (10)$$

The operator '$\arg\min_{\boldsymbol{\theta}}[\cdot]$' denotes the value of $\boldsymbol{\theta}$ that minimizes the NLLF in Eqn. (8). In particular, the mean $\widehat{\boldsymbol{\theta}}$ is known as the MAP estimate of modal parameters $\boldsymbol{\theta}$. The PCM $\widehat{\boldsymbol{\Sigma}}$ is the inverse of the Hessian matrix of the NLLF at the MAP, quantifying the remaining uncertainties of modal parameters. Note that the pseudoinverse (represented by '$[\cdot]\dagger$') is used here, because the Hessian matrix is rank-deficient due to the unit norm constraints on mode shapes. A fast way to calculate $\widehat{\boldsymbol{\theta}}$ and $\widehat{\boldsymbol{\Sigma}}$ is detailed in the following based on complex matrix calculus.

### 3.1 Maximum a posterior estimate

The MAP of modal parameters $\boldsymbol{\theta}$ is obtained by minimizing the NLLF. Brute-force numerical optimization is possible but inefficient, especially when handling large datasets from multi-setup vibration tests. Fortunately, the gradient of NLLF with respect to (w.r.t.) most of parameters can be derived analytically, enabling the use of the coordinate descent algorithm [38]. The coordinate descent algorithm begins with an initial guess of parameters, and iteratively updates one subset while fixing the remaining ones at their current estimates, until convergence is reached. To obtain the update, the first-order derivative of the NLLF w.r.t. each (vector) parameter is set to be zero, yielding the following analytical formula:

$$S_e^{(r)} = \frac{1}{d_r N_f^{(r)}} \sum_{k=1}^{N_f^{(r)}} \left(\widehat{Y}_k^{(r)} - S^{(r)}\boldsymbol{\Phi} H_k^{(r)} \boldsymbol{\Lambda}^{(\mathcal{S}(r))\mathrm{T}} \ddot{U}_k^{(r)}\right)^* \left(\widehat{Y}_k^{(r)} - S^{(r)}\boldsymbol{\Phi} H_k^{(r)} \boldsymbol{\Lambda}^{(\mathcal{S}(r))\mathrm{T}} \ddot{U}_k^{(r)}\right) \quad (11)$$

$$\mathrm{vec}^{\mathrm{T}}(\boldsymbol{\Phi}) = \sum_{r=1}^{n_r} S_e^{(r)-1} \mathrm{Re}\left[\mathrm{vec}^{\mathrm{T}}\left(\sum_{k=1}^{N_f^{(r)}} S^{(r)\mathrm{T}} \widehat{Y}_k^{(r)} \ddot{U}_k^{(r)*} \boldsymbol{\Lambda}^{(\mathcal{S}(r))} H_k^{(r)*}\right)\right] \cdot \left\{\sum_{r=1}^{n_r} S_e^{(r)-1}\left[\mathrm{Re}\left(\sum_{k=1}^{N_f^{(r)}} B_k^{(r)}\right) \otimes \left(S^{(r)\mathrm{T}} S^{(r)}\right)\right]\right\}^{-1} \quad (12)$$

$$\mathrm{vec}(\boldsymbol{\Lambda}^{(s)}) = \left\{\sum_{r\in\mathcal{R}(s)} S_e^{(r)-1} \mathrm{Re}\left[\sum_k [C_k^{(r)\mathrm{T}} \otimes (\ddot{U}_k^{(r)} \ddot{U}_k^{(r)*})]\right]\right\}^{-1} \left\{\sum_{r\in\mathcal{R}(s)} S_e^{(r)-1} \mathrm{Re}\left[\sum_{k=1}^{N_f^{(r)}} H_k^{(r)} \otimes (\ddot{U}_k^{(r)} \widehat{Y}_k^{(r)*})\right] \mathrm{vec}(S^{(r)}\boldsymbol{\Phi})\right\} \quad (13)$$

where we have defined $B_k^{(r)} = H_k^{(r)} \boldsymbol{\Lambda}^{(\mathcal{S}(r))\mathrm{T}} \ddot{U}_k^{(r)} \ddot{U}_k^{(r)*} \boldsymbol{\Lambda}^{(\mathcal{S}(r))} H_k^{(r)*}$ and $C_k^{(r)} = H_k^{(r)*} \boldsymbol{\Phi}^{\mathrm{T}} S^{(r)\mathrm{T}} S^{(r)} \boldsymbol{\Phi} H_k^{(r)}$. Here, $s = \mathcal{S}(r)$ defines a map from the $r$-th setup of sensors to the $s$-th setup of shakers, and $\mathcal{R}(s)$ defines a map from the $s$-th setup of shakers to all the corresponding setups of sensors. The detailed deviation of the above equations is postponed to Appendix A for reference. Due to the nonlinear coupling in the likelihood function, closed-form expressions for updating natural frequencies and damping ratios are unavailable. Since their dimension is typically low ($2m$, generally $m \leq 3$), the simplex search method [39] is used for their updates.

Since the coordinate descent algorithm is a local optimization method in nature, a proper initialization is crucial for a fast convergence to physically reasonable results. In the proposed method, initial guesses for natural frequencies, damping ratios, mode shapes and MPFs are need. Specifically, the damping ratio is initialized to be 1%, while the natural frequencies are estimated by identifying peaks from the singular value (SV) spectrum (i.e., singular values of estimated PSD matrix). For multi-setup problems, it is essential to consider multiple SV spectra from different setups to avoid misleading information due to improper configurations. For the initial mode shape, we first estimate the local mode shape as the left singular vector corresponding to the largest value of the estimated FRF matrix, and then assemble them into the global mode shape with a local least squares approach [12]. The estimation of initial local mode shape can be understood as follows. Assuming a single mode in the selected frequency band, the theoretical FRF at the initial (resonant) frequency $f_i$ reduces to $h_k^{(r)}(f_i) = \mathrm{i} S^{(r)} \boldsymbol{\phi}_i \boldsymbol{\lambda}_i^{(s)\mathrm{T}}/2\zeta_i$, which is a rank-one matrix with $S^{(r)}\boldsymbol{\phi}_i$ being its left singular vector and $\boldsymbol{\lambda}_i^{(s)}$ being its right singular vector corresponding to the sole nonzero singular value. Overall, the semi-analytical coordinate descent algorithm can reduce the computational complexity, making it effective for the MAP estimation of modal parameters.

### 3.2 Posterior covariance matrix calculation

In the context of Laplace approximation, the remaining uncertainty in the identified modal parameters is quantified by the PCM, which is equal to the inverse of Hessian matrix of NLLF at the MAP. A free parameter mapping-based approach

[13] is adopted to calculate the PCM as

$$[\nabla^2 \hat{\mathcal{L}}(\boldsymbol{\theta})]^\dagger = \nabla \hat{\boldsymbol{v}}_c(\boldsymbol{w})[\nabla^2 \hat{L}_c(\boldsymbol{w})]^\dagger \nabla^T \hat{\boldsymbol{v}}_c(\boldsymbol{w}) \quad (14)$$

where $\boldsymbol{w} \in \mathbb{R}^p$ is a set of free parameters that automatically satisfies equality constraints $\boldsymbol{G}(\boldsymbol{\theta}) = [G_1(\boldsymbol{\theta}), G_2(\boldsymbol{\theta}), \dots, G_m(\boldsymbol{\theta})]^T = \boldsymbol{0}$; $\boldsymbol{v}_c(\boldsymbol{w})$ defines a mapping from $\boldsymbol{w}$ to $\boldsymbol{\theta}$ and $L_c(\boldsymbol{w}) = L(\boldsymbol{v}_c(\boldsymbol{w}))$. The symbol '$\widehat{[\cdot]}$' indicates the PCM is evaluated at the MAP, i.e., $\boldsymbol{\theta} = \widehat{\boldsymbol{\theta}}$. The Hessian matrix of $L_c(\boldsymbol{w})$ is expressed as [13]

$$\nabla^2 \hat{L}_c(\boldsymbol{w}) = \nabla^T \hat{\boldsymbol{v}}_c(\boldsymbol{w}) \nabla^2 \hat{\mathcal{L}}(\boldsymbol{\theta}) \nabla \hat{\boldsymbol{v}}_c(\boldsymbol{w}) + \left(\boldsymbol{I}_p \otimes \nabla \hat{\mathcal{L}}(\boldsymbol{\theta})\right) \nabla^2 \hat{\boldsymbol{v}}_c(\boldsymbol{w}) \quad (15)$$

where '$\otimes$' denotes the Kronecker product; $\boldsymbol{I}_p \in \mathbb{R}^{p \times p}$ denotes a $p$-by-$p$ identity matrix. In constrained optimization problems, the gradient is typically not zero at the optimum because the stationary point may not be feasible for given constraints. Luckily, for the specific problem considered here, $\nabla \hat{L}(\boldsymbol{w})$ is indeed zero as proofed in Ref. [13], and hence the second term in Eqn. (15) can be ignored legitimately. The PCM in Eqn. (14) remains the same no matter what admissible $\boldsymbol{v}_c(\boldsymbol{w})$ and $\boldsymbol{w}$ are chosen. As informed by Eqn. (15), the key question lies in the choice of Jacobian matrix $\nabla \hat{\boldsymbol{v}}_c(\boldsymbol{w})$. According to Ref. [13], we choose $\nabla \hat{\boldsymbol{v}}_c(\boldsymbol{w})$ to be the matrix consisting of basis vectors spanning the nullspace of the Jacobian matrix $\nabla \hat{\boldsymbol{g}}(\boldsymbol{\theta})$, i.e., $\nabla \hat{\boldsymbol{v}}_c(\boldsymbol{w}) = \text{null}(\nabla \hat{\boldsymbol{g}}(\boldsymbol{\theta})) \in \mathbb{R}^{n_\theta \times (n_\theta - m)}$. Substituting it into Eqn. (15) and subsequently into Eqn. (14), one has

$$\widehat{\boldsymbol{\Sigma}} = \nabla \hat{\boldsymbol{v}}_c(\boldsymbol{w})[\nabla \hat{\boldsymbol{v}}_c(\boldsymbol{w})^T \nabla^2 \hat{\mathcal{L}}(\boldsymbol{\theta}) \nabla \hat{\boldsymbol{v}}_c(\boldsymbol{w})]^{-1} \nabla \hat{\boldsymbol{v}}_c(\boldsymbol{w})^T \quad (16)$$

Note that the pseudo-inverse operation is replaced by the conventional inverse, because the matrix within the brackets is of full rank, removing potential numerical errors associated with the matrix pseudo-inverse.

In order to obtain the PCM $\widehat{\boldsymbol{\Sigma}}$ in Eqn. (16), the remaining issue is to derive the Hessian matrix $\nabla^2 \hat{\mathcal{L}}(\boldsymbol{\theta})$. Here, we derive it using complex matrix calculus. For clarity, we define $\mathcal{L}^{(ab)} = \partial^2 \mathcal{L}(\boldsymbol{\theta})/\partial a \partial b$, where '$a$' and '$b$' represents the parameters in $\boldsymbol{\theta}$. The second-order derivative of NLLF w.r.t. all unknown parameters are presented in Table 1 and Table 2, with detailed derivations postponed in Appendix B.

**TABLE 1.** Expressions of Hessian matrix of NLLF

| Notation | Results |
|---|---|
| $\mathcal{L}^{(\theta_m \theta_m)}$ | $\sum_{r=1}^{n_r} \sum_{k=1}^{N_f^{(r)}} 2 S_e^{(r)-1} \text{Re}\left\{ L_{k1}^{(r)} \boldsymbol{L}_d^T \left[ \left( \boldsymbol{\Lambda}^{(S(r))T} \ddot{\boldsymbol{U}}_k^{(r)} \ddot{\boldsymbol{U}}_k^{(r)*} \boldsymbol{\Lambda}^{(S(r))} \right) \otimes \left( \boldsymbol{\Phi}^T \boldsymbol{S}^{(r)T} \boldsymbol{S}^{(r)} \boldsymbol{\Phi} \right) \right] \boldsymbol{L}_d \bar{L}_{k1}^{(r)T} \right\} + \sum_{r=1}^{n_r} \sum_{k=1}^{N_f^{(r)}} L_{k5}^{(r)}$ |
| $\mathcal{L}^{(\theta_m \Phi)}$ | $\sum_{r=1}^{n_r} \sum_{k=1}^{N_f^{(r)}} \left\{ [L_{k1}^{(r)} \ \bar{L}_{k1}^{(r)}] \cdot (\boldsymbol{I}_2 \otimes \boldsymbol{L}_d^T) \cdot \begin{bmatrix} L_{k2}^{(r)} \\ \bar{L}_{k2}^{(r)} \end{bmatrix} \right\}$ |
| $\mathcal{L}^{(\theta_m \Lambda^{(s)})}$ | $\sum_{r \in \mathcal{R}(s)} \sum_{k=1}^{N_f^{(r)}} \left\{ [L_{k1}^{(r)} \ \bar{L}_{k1}^{(r)}] \cdot (\boldsymbol{I}_2 \otimes \boldsymbol{L}_d^T) \cdot \begin{bmatrix} L_{k3}^{(r)} \\ \bar{L}_{k3}^{(r)} \end{bmatrix} \right\}$ |
| $\mathcal{L}^{(\theta_m S_e^{(r)})}$ | $\sum_{k=1}^{N_f^{(r)}} \left\{ [L_{k1}^{(r)} \ \bar{L}_{k1}^{(r)}] \cdot (\boldsymbol{I}_2 \otimes \boldsymbol{L}_d^T) \cdot \begin{bmatrix} L_{k4}^{(r)} \\ \bar{L}_{k4}^{(r)} \end{bmatrix} \right\}$ |
| $\mathcal{L}^{(\Phi \Phi)}$ | $\sum_{r=1}^{n_r} 2 S_e^{(r)-1} \left[ \text{Re}\left( \sum_{k=1}^{N_f^{(r)}} \boldsymbol{B}_k^{(r)} \right) \otimes \left( \boldsymbol{S}^{(r)T} \boldsymbol{S}^{(r)} \right) \right]$ |
| $\mathcal{L}^{(\Phi \Lambda^{(s)})}$ | $\sum_{r \in \mathcal{R}(s)} \left\{ -2S_e^{(r)^{-1}} \sum_{k=1}^{N_f^{(r)}} \text{Re}[\overline{H}_k^{(r)} \otimes (S^{(r)\text{T}} \widehat{Y}_k^{(r)} \ddot{U}_k^{(r)*})] + 2S_e^{(r)^{-1}} \sum_{k=1}^{N_f^{(r)}} [\overline{H}_k^{(r)} \otimes (S^{(r)\text{T}} S^{(r)} \Phi H_k^{(r)} \Lambda^{(S(r))\text{T}} \ddot{U}_k^{(r)} \ddot{U}_k^{(r)*})] \right.$ $\left. + 2S_e^{(r)^{-1}} \sum_{k=1}^{N_f^{(r)}} \left[ (\overline{H}_k^{(r)} \Lambda^{(S(r))\text{T}} \overline{\ddot{U}_k^{(r)}} \ddot{U}_k^{(r)*}) \otimes (S^{(r)\text{T}} S^{(r)} \Phi H^{(r)}) \right] K_{ml} \right\}$ |
| $\mathcal{L}^{(\Phi S_e^{(r)})}$ | $2S_e^{(r)^{-2}} \text{Re} \left[ \text{vec} \left( \sum_{k=1}^{N_f^{(r)}} S^{(r)\text{T}} \widehat{Y}_k^{(r)} \ddot{U}_k^{(r)*} \Lambda^{(S(r))} H_k^{(r)*} \right) \right] - 2S_e^{(r)^{-2}} \left[ \text{Re} \left( \sum_{k=1}^{N_f^{(r)}} B_k^{(r)} \right) \otimes (S^{(r)\text{T}} S^{(r)}) \right] \text{vec}(\Phi)$ |
| $\mathcal{L}^{(\Lambda^{(s)} \Lambda^{(s)})}$ | $\sum_{r \in \mathcal{R}(s)} 2S_e^{(r)^{-1}} \left[ \text{Re} \left( \sum_{k=1}^{N_f^{(r)}} C_k^{(r)\text{T}} \right) \otimes (\ddot{U}_k^{(r)} \ddot{U}_k^{(r)*}) \right]$ |
| $\mathcal{L}^{(\Lambda^{(s)} S_e^{(r)})}$ | $2S_e^{(r)^{-2}} \text{Re} \left[ \text{vec} \left( \sum_{k=1}^{N_f^{(r)}} \ddot{U}_k^{(r)} \widehat{Y}_k^{(r)*} S^{(r)} \Phi H_k^{(r)} \right) \right] - 2S_e^{(r)^{-2}} \left[ \text{Re} \sum_{k=1}^{N_f^{(r)}} C_k^{(r)\text{T}} \otimes (\ddot{U}_k^{(r)} \ddot{U}_k^{(r)*}) \right] \text{vec}(\Lambda^{(S(r))})$ |
| $\mathcal{L}^{(S_e^{(r)} S_e^{(r)})}$ | $-d_r N_f^{(r)} S_e^{(r)^{-2}} + 2S_e^{(r)^{-3}} \sum_{k=1}^{N_f^{(r)}} A_k^{(r)}$ |

Note: $\boldsymbol{\theta}_m = [f_1, \zeta_1, \dots f_m, \zeta_m]^{\text{T}}$; $L_{kj}^{(r)}$ is listed in Table 2 for $j = 1, \dots, 5$; $L_{kj}^{(r)}$ and $\overline{L}_{kj}^{(r)}$ are conjugate pairs; $L_d \in \mathbb{R}^{m^2 \times m}$ is a commutation matrix; $A_k^{(r)} = \left[ \widehat{Y}_k^{(r)} - S^{(r)} \Phi H_k^{(r)} \Lambda^{(s)\text{T}} \ddot{U}_k^{(r)} \right]^* \left[ \widehat{Y}_k^{(r)} - S^{(r)} \Phi H_k^{(r)} \Lambda^{(s)\text{T}} \ddot{U}_k^{(r)} \right]$

**TABLE 2.** Companion expression for Hessian of NLLF

| Notation | Results |
|---|---|
| $L_{k1}^{(r)}$ | $\begin{bmatrix} \frac{\partial h_{1k}^{(r)}}{\partial f_1} & \frac{\partial h_{1k}^{(r)}}{\partial \zeta_1} & \cdots & 0 & 0 \\ \vdots & \vdots & \ddots & \vdots & \vdots \\ 0 & 0 & \cdots & \frac{\partial h_{1k}^{(r)}}{\partial f_m} & \frac{\partial h_{1k}^{(r)}}{\partial \zeta_m} \end{bmatrix}^{\text{T}}$ |
| $L_{k2}^{(r)}$ | $-S_e^{(r)^{-1}} (\Lambda^{(S(r))\text{T}} \ddot{U}_k^{(r)} \widehat{Y}_k^{(r)*} S^{(r)} \otimes I_m) K_{md} + S_e^{(r)^{-1}} (\Lambda^{(S(r))\text{T}} \ddot{U}_k^{(r)} \ddot{U}_k^{(r)*} \Lambda^{(S(r))} \overline{H}_k^{(r)}) \otimes (\Phi^{\text{T}} S^{(r)\text{T}})$ $+ S_e^{(r)^{-1}} [(\Lambda^{(S(r))\text{T}} \ddot{U}_k^{(r)} \ddot{U}_k^{(r)*} \Lambda^{(S(r))} \overline{H}_k^{(r)} \Phi^{\text{T}} S^{(r)\text{T}} S^{(r)}) \otimes I_m] K_{md}$ |
| $L_{k3}^{(r)}$ | $-S_e^{(r)^{-1}} \left[ I_m \otimes \left( \Phi^{\text{T}} S^{(r)\text{T}} \overline{Y_k^{(r)} \ddot{U}_k^{(r)*}} \right) \right] + S_e^{(r)^{-1}} \left[ I_m \otimes \left( \Phi^{\text{T}} S^{(r)\text{T}} S^{(r)} \Phi \overline{H}_k^{(r)} \Lambda^{(S(r))\text{T}} \overline{\ddot{U}_k^{(r)} \ddot{U}_k^{(r)*}} \right) \right]$ $+ S_e^{(r)^{-1}} [(\Lambda^{(S(r))\text{T}} \ddot{U}_k^{(r)} \ddot{U}_k^{(r)*}) \otimes (\Phi^{\text{T}} S^{(r)\text{T}} S^{(r)} \Phi \overline{H}_k^{(r)})] K_{ml}$ |
| $L_{k4}^{(r)}$ | $S_e^{(r)^{-2}} \text{vec} \left[ \Phi^{\text{T}} S^{(r)\text{T}} \left( \overline{Y}_k^{(r)} \ddot{U}_k^{(r)*} - S^{(r)} \Phi \overline{H}_k^{(r)} \Lambda^{(S(r))\text{T}} \overline{\ddot{U}_k^{(r)} \ddot{U}_k^{(r)*}} \right) \Lambda^{(S(r))} \right]$ |
| $L_{k5}^{(r)}$ | $\text{blkdiag} \left\{ \begin{matrix} L_{k6}^{(\theta_m \theta_m)} \cdot \begin{bmatrix} [\partial^2 h_{1k}^{(r)}/\partial f_i \partial f_i, \dots, \partial^2 h_{mk}^{(r)}/\partial f_i \partial f_i]^{\text{T}} \\ [\partial^2 \overline{h}_{1k}^{(r)}/\partial f_i \partial f_i, \dots, \partial^2 \overline{h}_{mk}^{(r)}/\partial f_i \partial f_i]^{\text{T}} \end{bmatrix} & L_{k6}^{(\theta_m \theta_m)} \cdot \begin{bmatrix} [\partial^2 h_{1k}^{(r)}/\partial f_i \partial \zeta_i, \dots, \partial^2 h_{mk}^{(r)}/\partial f_i \partial \zeta_i]^{\text{T}} \\ [\partial^2 \overline{h}_{1k}^{(r)}/\partial f_i \partial \zeta_i, \dots, \partial^2 \overline{h}_{mk}^{(r)}/\partial f_i \partial \zeta_i]^{\text{T}} \end{bmatrix} \\ L_{k6}^{(\theta_m \theta_m)} \cdot \begin{bmatrix} [\partial^2 h_{1k}^{(r)}/\partial \zeta_i \partial f_i, \dots, \partial^2 h_{mk}^{(r)}/\partial \zeta_i \partial f_i]^{\text{T}} \\ [\partial^2 \overline{h}_{1k}^{(r)}/\partial \zeta_i \partial f_i, \dots, \partial^2 \overline{h}_{mk}^{(r)}/\partial \zeta_i \partial f_i]^{\text{T}} \end{bmatrix} & L_{k6}^{(\theta_m \theta_m)} \cdot \begin{bmatrix} [\partial^2 h_{1k}^{(r)}/\partial \zeta_i \partial \zeta_i, \dots, \partial^2 h_{mk}^{(r)}/\partial \zeta_i \partial \zeta_i]^{\text{T}} \\ [\partial^2 \overline{h}_{1k}^{(r)}/\partial \zeta_i \partial \zeta_i, \dots, \partial^2 \overline{h}_{mk}^{(r)}/\partial \zeta_i \partial \zeta_i]^{\text{T}} \end{bmatrix} \end{matrix} \right\}$ for $i = 1, \dots, m$ |
| $L_{k6}^{(r)}$ | $[L_{k7}^{(r)} \ \overline{L}_{k7}^{(r)}] \cdot (I_2 \otimes L_d)$ |
| $L_{k7}^{(r)}$ | $S_e^{(r)^{-1}} \text{vec}^{\text{T}} \left[ -\Phi^{\text{T}} S^{(r)\text{T}} \left( \overline{Y}_k^{(r)} \ddot{U}_k^{(r)*} - S^{(r)} \Phi \overline{H}_k^{(r)} \Lambda^{(S(r))\text{T}} \overline{\ddot{U}_k^{(r)} \ddot{U}_k^{(r)*}} \right) \Lambda^{(S(r))} \right]$ |

Note: The derivatives of $h_k$ w.r.t. $f$ and $\zeta$ are listed at Appendix C; "bkldiag{·}" denotes a block diagonal matrix.

To enhance readability, the pseudo-code of the proposed modal identification algorithm is presented in Algorithm 1. First, we obtain the (scaled) FFTs of the measured structural input and output vibration data, i.e., $\{\hat{\boldsymbol{Y}}_k^{(r)}\}_{r=1}^{n_r}$ and $\{\ddot{\boldsymbol{U}}_k^{(r)}\}_{r=1}^{n_r}$. Next, based on the SV spectra of estimated PSD matrix, a frequency band is chosen, along with initial guesses for the natural frequencies, damping ratios and mode shapes within this band. Iteration of coordinate descent then starts using the analytical updating in Eqns. (11)- (13). The convergence is achieved when the NLLF cannot be further reduced or the maximum number of iterations is reached. The PCM of the modal parameters is finally evaluated with the expressions given in Table. 1 and Table. 2.

---

**Algorithm 1.** Pseudo-code for MAP and PCM computation in the Bayesian FFT method for multi-setup EMA

**Input:** Measured structural input data $\{\ddot{\boldsymbol{U}}_k^{(r)}\}_{r=1}^{n_r}$ and output data $\{\hat{\boldsymbol{Y}}_k^{(r)}\}_{r=1}^{n_r}$;

Number of test setups $n_r$, Number of shaker placement schemes $n_s$, Selection matrices $\{\boldsymbol{S}^{(r)}\}_{r=1}^{n_r}$, Mapping $\mathcal{S}(r)$;

Authorized convergence criterion $\epsilon_0$, Maximum number of iterations $Itermax$;

**Output:** MAP of modal parameters $\hat{\boldsymbol{\theta}}$; Posterior covariance matrix $\hat{\boldsymbol{\Sigma}}$;

% Part I: MAP calculation.

1. Set initial values of $\boldsymbol{\theta}_{m(0)}$, $\boldsymbol{\Phi}_{(0)}$ and $\{\boldsymbol{\Lambda}_{(0)}^{(s)}\}_{s=1}^{n_s}$;    % see Section 3.1
2. Set $t = 0$;
3. **While** $t < Itermax$    % authorized maximum number of iterations, default 100
4.    Update $\{S_{e(t)}^{(r)}\}_{r=1}^{n_r}$;    % see Eqn.(11)
5.    Set $\hat{\boldsymbol{\theta}} = \{\boldsymbol{\theta}_{m(t)}, \boldsymbol{\Phi}_{(t)}, \{S_{e(t)}^{(r)}\}_{r=1}^{n_r}, \{\boldsymbol{\Lambda}_{(t)}^{(s)}\}_{s=1}^{n_s}\}$;
6.    Calculate $\mathcal{L}_{(t)}(\hat{\boldsymbol{\theta}}) = \mathcal{L}(\hat{\boldsymbol{\theta}})$;    % see Eqn.(8)
7.    **If** $t > 0$
8.      $\epsilon = (\mathcal{L}_{(t-1)} - \mathcal{L}_{(t)})/\mathcal{L}_{(t-1)}$;
9.      **If** $\epsilon < \epsilon_0$    % authorized convergence criterion, default $\epsilon_0 = 10^{-6}$
10.        STOP;    % convergence achieved
11.      **End If**
12.    **End If**
13.    Set $t = t + 1$;
14.    Update $\{\boldsymbol{\Lambda}_{(t)}^{(s)}\}_{s=1}^{n_s}$;    % see Eqn.(13)
15.    Update $\boldsymbol{\Phi}_{(t)}$;    % see Eqn.(12)
16.    Re-normalize $\{\boldsymbol{\Lambda}_{(t)}^{(s)}\}_{s=1}^{n_s}$ and $\boldsymbol{\Phi}_{(t)}$ to ensure mode shape unit-norm constraint;
17.    Update $\boldsymbol{\theta}_{m(t)}$ using Nelder-Mead Simplex Method;
18. **End While**

% Part II: PCM evaluation

21. Calculate $\mathcal{L}^{(ab)}$ and assemble it into Hessian matrix $\frac{\partial^2 \mathcal{L}(\hat{\boldsymbol{\theta}})}{\partial \boldsymbol{\theta} \partial \boldsymbol{\theta}^{\mathrm{T}}}$;    % see Table. 1 and Table. 2
22. Solve $\nabla \hat{\boldsymbol{v}}_c(\boldsymbol{w}) = \mathrm{null}(\nabla \hat{\boldsymbol{g}}(\hat{\boldsymbol{\theta}}))$ to get $\nabla \hat{\boldsymbol{v}}_c(\boldsymbol{w})$;    % see Section 3.2
23. Calculate the posterior covariance matrix $\hat{\boldsymbol{\Sigma}}$    % see Eqn.(15)

# 4    Illustrative examples

Empirical studies with synthetic data and field test data from bridge and building structures are presented to evaluate the consistency and performance of the proposed algorithm. The method is able to deal with cases of multi-directional excitations. However, the following example will use only a uni-axial shaker and several sensors, representing a type of low-cost forced vibration test. Specifically, the proposed algorithm accommodates multi-directional excitations using a uni-axial shaker and large-scale measurements with a limited number of sensors through a multi-setup strategy. These examples also serve to provide guidance for the application of the proposed method. For simplicity, we will refer to the proposed method as MS-BAYEMA (Multi-setup Bayesian Experimental Modal Analysis) hereafter.

## 4.1    Verification with synthetic data

This section presents two numerical examples based on an 18-m simply-supported bridge and a six-story shear-type building. Since the true modal parameters are known a priori, it is possible to use the numerical examples to verify the consistency of the proposed algorithm.

### 4.1.1    An 18-m simply-supported bridge

Consider an idealized 18-m simply-supported bridge, as depicted in Fig 1, with a total of 20 potential measured DOFs in the Y- and Z-directions. The modal properties of the structure are given in Table 3. Notably, the third and fourth modes have mode shapes coupled in two orthogonal directions, a phenomenon often observed in long bridges [21]. For this forced vibration test, we assume the availability of seven bi-axial accelerometers and a uni-axial shaker. A straightforward and efficient test configuration is illustrated in Fig. 1, where two accelerometers serve as reference sensors and the remaining four accelerometers act as rover sensors to measure the structural responses in each setup. This test configuration encompasses two setups to cover all 20 DOFs by moving the sensors and additional two setups to excite the modes from two directions by moving the shaker, resulting in a four-setup single-source forced vibration test. The shaker location, reference sensors location, and rover sensors location in each setup are represented by a red star, blue circles, and grey rectangles in Fig. 1, respectively.

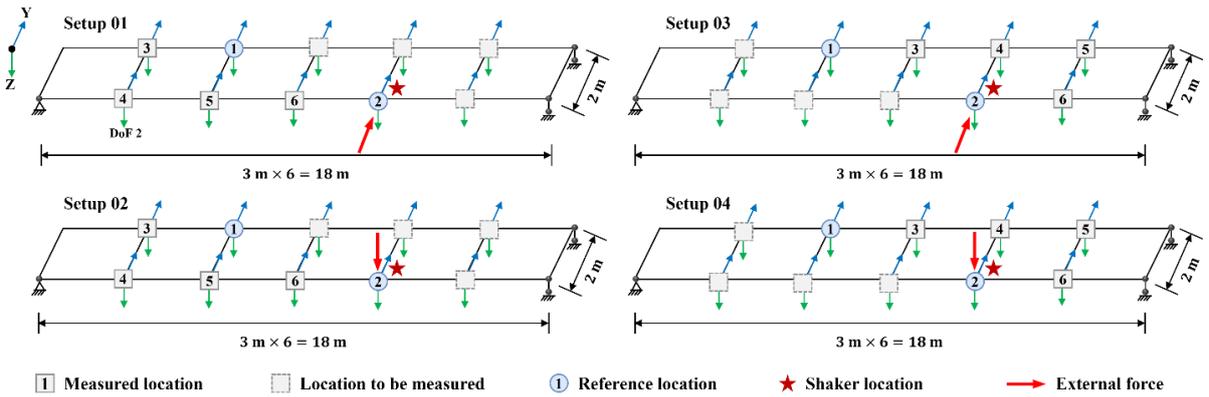

Fig. 1  Geometric property of the simply-supported bridge and test configuration, synthetic bridge data

It is assumed that the shaker generates a pseudo-random excitation with a flat root PSD over the frequency band from 0.1 Hz to 10 Hz. Synthetic structural acceleration responses under this excitation are simulated using the modal superposition rule. The structural input and output are assumed to be measured using commercial piezoelectric accelerometers, which introduce measurement noise modeled as a Gaussian white noise process with a constant PSD of $S_e = 1 \times 10^{-5}$ g/$\sqrt{\text{Hz}}$ . The raw data, acquired at a sampling frequency of 100 Hz, spans a total duration of 70 seconds for each setup. Specifically, the data includes 5 seconds of random vibration before the shaker is turned on, 60 seconds of forced vibration during the shaker is turned on and 5 seconds of free vibration after the shaker is turned off. The structural input measured during setup 01 is shown in Fig. 2 (a) and (b) in the manner of time history and root PSD, respectively. A typical structural response at DOF 2 is depicted in Fig. 2 (c). The SV spectrum calculated using the data from Setup 01 and Setup 02 is shown in Fig. 2 (d), where clear spectral peaks indicate the existence of four modes. They are further selected as two pairs of well-separated modes and one pair of closely-spaced modes for modal identification. Resonance bands and initial guesses of natural frequency are depicted with brackets and circles in Fig 2(d), respectively.

Table 3 True modal parameter values, synthetic bridge data

| Modal parameters | | Setting values |
|---|---|---|
| Frequency $f$ (Hz) | | $[1.22, 4.74, 5.76, 5.89]^T$ |
| Damping ratio $\zeta$ (%) | | $[2.00, 2.00, 2.00, 2.00]^T$ |
| Mode shape $\phi_1$ | Y-Dir. | $[0, 0, 0, 0, 0, 0, 0, 0, 0, 0]^T$ |
| | Z-Dir. | $[-0.20, -0.35, -0.41, -0.35, -0.20, -0.20, -0.35, -0.41, -0.35, -0.20]^T$ |
| Mode shape $\phi_2$ | Y-Dir. | $[0, 0, 0, 0, 0, 0, 0, 0, 0, 0]^T$ |
| | Z-Dir. | $[-0.35, -0.35, 0, 0.35, -0.35, -0.35, -0.35, 0, 0.35, 0.35]^T$ |
| Mode shape $\phi_3$ | Y-Dir. | $[0.19, 0.36, 0.42, 0.36, 0.19, 0.19, 0.36, 0.42, 0.36, 0.19]^T$ |
| | Z-Dir. | $[-0.06, -0.11, -0.12, -0.11, -0.06, 0.06, 0.11, 0.12, 0.11, 0.06]^T$ |
| Mode shape $\phi_4$ | Y-Dir. | $[0.06, 0.11, 0.13, 0.11, 0.06, 0.06, 0.11, 0.13, 0.11, 0.06]^T$ |
| | Z-Dir. | $[-0.20, -0.35, -0.41, -0.35, -0.20, 0.20, 0.35, 0.41, 0.35, 0.20]^T$ |

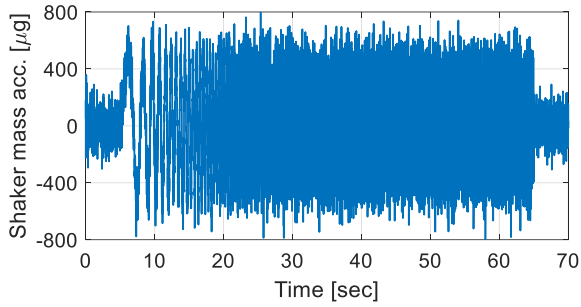
(a) Time history of shaker acceleration; Setup 01

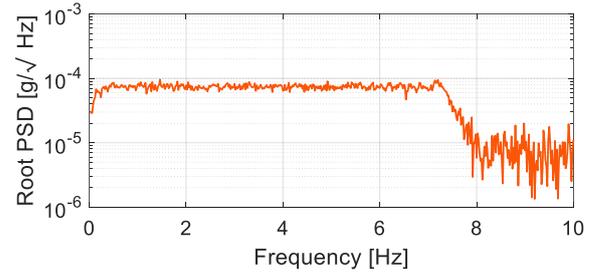
(b) Root PSD of shaker acceleration; Setup 01

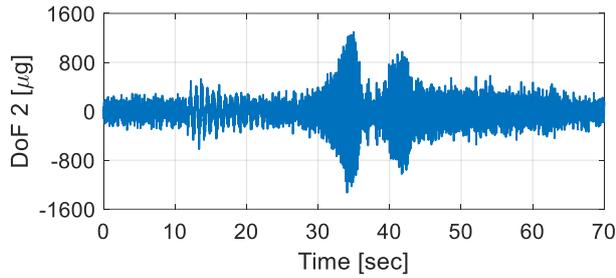
(c) Time history of structural response at DOF 2; Setup01

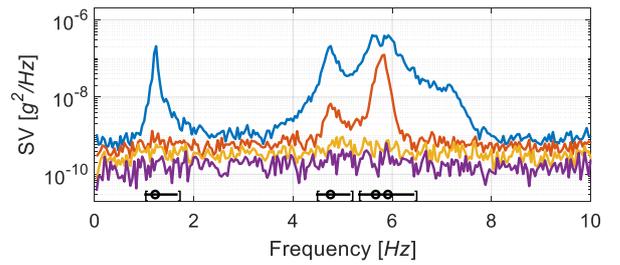
(d) SV spectrum; Setup01 and Setup02

Fig. 2 Typical measured structural input and output, synthetic bridge data

The SIMO vibration data in the selected frequency bands are provided to MS-BAYEMA for modal identification. The identified modal parameters of the four targeted modes are shown in Fig. 3, with the corresponding identified natural frequency, damping ratio and modal assurance criterion (MAC). The MAC, defined as the cosine of the hyper-angle between two mode shape vectors, is adopted for quantifying the discrepancy between them. The identified mode shapes are also depicted in Fig. 3, where the red and black lines represent the undeformed and deformed shapes of the bridge, respectively. The close alignment of the identified results with the nominal values and the high MAC scores validates the effectiveness of the developed algorithm. In addition to MAP, the MS-BAYEMA algorithm quantifies identification uncertainty through PCM. As shown in Fig. 3, the values in parentheses indicate the coefficient of variation (c.o.v. = standard derivation/mean) of identified modal parameters. Since the mode shape is a random vector, its c.o.v. is defined as the square root of the trace of the PCM [9]. Notably, a low level of identification uncertainty is achieved for all parameters. Furthermore, the identified results of MPF are presented in Table 4. The results demonstrate that the MS-BAYEMA provides well-identified values of MPF, as evidenced by the agreement between the identified values and the true values.

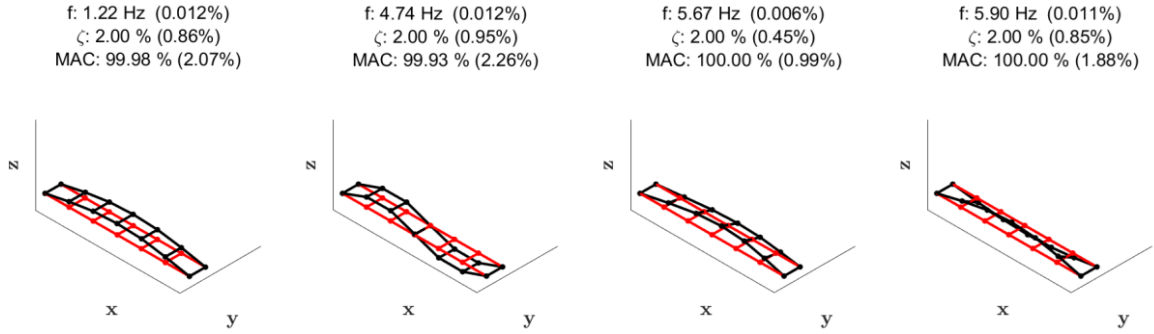

Fig. 3  Identified results for four modes; synthetic bridge data

Table 4  Identified results of MPF, synthetic bridge data

| Mode | | 1 | | 2 | | 3 | | 4 | |
|---|---|---|---|---|---|---|---|---|---|
| | | True | Identified | True | Identified | True | Identified | True | Identified |
| MPF | Y-direction | 0.0000 | 0.0000 | 0.0000 | 0.0000 | 0.0037 | 0.0037 (1.88%) | 0.0011 | 0.0011 (0.38%) |
| | Z-direction | 0.0035 | 0.0035 (0.63%) | 0.0035 | 0.0035 (0.75%) | 0.0011 | 0.0011 (0.72%) | 0.0037 | 0.0037 (1.06%) |

Note: The values in parentheses indicate the c.o.v.s.

In the following illustrative scenario, the algorithm in Ref. [31] for single-setup forced vibration test with single-input ('referred to SS-BAYEMA' hereafter) is included for comparison with the proposed MS-BAYEMA algorithm. Three different cases are considered, labeled SSB1, SSB2, and SSB3 respectively. Case SSB1 is the previous multi-setup vibration test shown in Fig. 1. Case SSB2 and SSB3 (shown in Fig. 4) represent single-setup vibration tests using 11 uni-axial accelerometers, with one accelerometer recording the structural input and the others measuring the structural output. The only distinction between cases SSB2 and SSB3 lies in the excitation DOF. Specifically, SSB2 targets at modes in Z-direction, while SSB3 targets at modes in Y-direction. The test duration for both single-setup tests is set to be twice that of the multi-setup test, i.e., 140 seconds, to have a consistent duration along each direction across all three cases for a fair comparison [16]. Throughout all three cases, the structural input, measurement noise, selected frequency bands, and initial guesses for natural frequency and damping ratio remain the same.

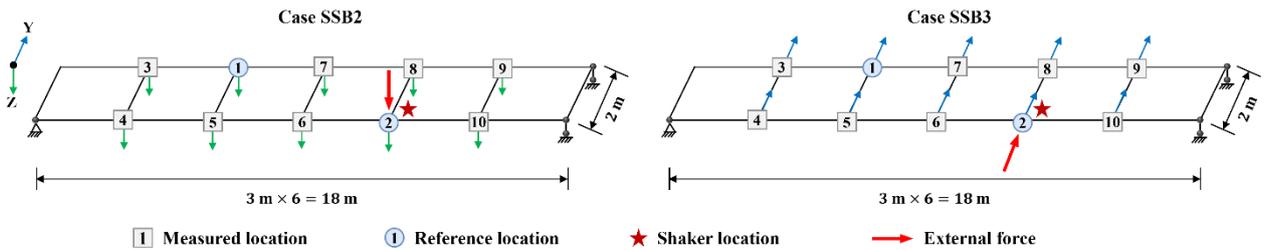

Fig. 4  Optional test configuration for single-setup vibration test, synthetic bridge data

The selected four modes are further identified by SS-BAYEMA and listed in Table 5. It should be noted that Case SSB2 accounts for vertical modes, while Case SSB3 accounts for transverse modes, rendering modes from the other direction unidentifiable. While all modal parameters identified by MS-BAYEMA closely resemble the true values, results obtained from SS-BAYEMA exhibit significant discrepancies for the third and fourth modes, which are coupled in the Y- and Z-directions. This deficiency arises from the fact that SS-BAYEMA uses structural input and output data collected from either Y- or Z-direction, thereby disregarding potential coupled vibrations from the other direction. This assumption introduces biases in cases where the structure exhibits coupled mode shapes in two orthogonal directions. Moreover, it is noted that the identification uncertainty for SS-BAYEMA is consistently smaller than that for MS-BAYEMA. This difference can be attributed to the incorporation of more vibration data in MS-BAYEMA. As investigated in Ref. [13], identifying closely-spaced or coupled modes in two orthogonal directions separately can introduce bias and underestimate the identification uncertainty. In this context, MS-BAYEMA effectively captures the coupling effects of mode shapes and the uncertainty arising due to the proximity of structural modes.

Table 5 identified modal parameters from synthetic data, synthetic bridge data

| Parameter | Mode | SSB1 (MS-BAYEMA) | SSB2 (SS-BAYEMA) | SSB3 (SS-BAYEMA) | Nominal values |
|---|---|---|---|---|---|
| Natural frequency $f_i$ (Hz) | 1 | 1.22 (0.012%) | 1.22 (0.009%) | N/A | 1.22 |
| | 2 | 4.74 (0.012%) | 4.74 (0.011%) | N/A | 4.74 |
| | 3 | 5.67 (0.006%) | N/A | 5.67 (0.005%) | 5.67 |
| | 4 | 5.90 (0.011%) | 5.89 (0.13%) | N/A | 5.90 |
| Damping ratio $\zeta_i$ (%) | 1 | 2.0 (0.86%) | 2.0 (0.65%) | N/A | 2.0 |
| | 2 | 2.0 (0.95%) | 2.0 (0.87%) | N/A | 2.0 |
| | 3 | 2.0 (0.45%) | N/A | 2.2 (0.56%) | 2.0 |
| | 4 | 2.0 (0.85%) | 2.2 (0.95%) | N/A | 2.0 |
| Mode shape MAC (%) | 1 | 99.98 (2.07%) | 99.99 (0.98%) | N/A | - |
| | 2 | 99.93 (2.26%) | 99.96 (1.29%) | N/A | - |
| | 3 | 100.00 (0.99%) | N/A | 95.78 (0.53%) | - |
| | 4 | 100.00 (1.88%) | 95.78 (1.38%) | N/A | - |

Note: 'N/A' indicates that the modes are not identified;
Values in parenthesis represent the c.o.v. of the corresponding modal parameters;

### 4.1.2 A six-story shear-type building

To further verify the performance of our algorithm for building structures, the modal identification of a lab-scale six-story shear-type building is considered, as depicted in Fig 5. The considered building model has a height of 1200 mm with each floor measuring 450 mm × 400 mm. Assuming a rigid floor, the floor plan is modeled by four beams with large enough stiffness. Details of the material and corresponding modal properties are given in Table 6. For this example, sensors are placed at four corners of each floor to record structural responses in the X- and Y-directions, resulting in a total of 48 DOFs to be measured. We assume that nine bi-axial accelerometers and a single-source shaker are available for the forced vibration test. The adopted test configuration is illustrated in Fig. 5, where one accelerometer records the external force, four accelerometers serve as reference sensors and the remaining ones serve as rover sensors to record the structural responses in each setup. Translational modes in both X- and Y-directions are of interest, so that the single-source shaker is placed diagonally at the floor to excite modes in the two orthogonal directions simultaneously. This test configuration encompasses five setups to cover 48 DOFs by solely moving rover sensors. The shaker location, reference sensors location, and rover sensors location in each setup are represented by a red star, blue circles, and grey rectangles in Fig. 5, respectively.

It is assumed that a pseudo-random excitation, characterized by a flat root PSD within the frequency range of 0.1 Hz to 15 Hz, is generated by the shaker and applied to the structure in the direction of red narrow, as shown in Fig 6(a) and Fig 6(b). Synthetic structural acceleration responses are simulated for a duration of 70 seconds at a sampling frequency of 100 Hz, followin the principle of mode superposition. This duration includes 5 seconds of random vibration before the shaker is turned on, 60 seconds of forced vibration and 5 seconds of free vibration after the shaker is turned off. The structural input and output are assumed to be measured with an added channel measurement noise with a constant PSD $S_e = 1 \times 10^{-5}$ g/$\sqrt{\text{Hz}}$. The measured structural responses in one channel are depicted in Fig 6(c). As shown in Fig 6(d), the SV spectrum plots six peaks, indicating six potential structural modes. To investigate the proposed method for identifying different numbers of modes within the selected band, the first three modes are identified together in the same band ($m = 3$), while the fourth and fifth modes are identified in a single band ($m = 2$) and the last mode is identified individually ($m = 1$). The (hand-picked) frequency bands and the initial guesses of natural frequency are shown with the brackets and circles, respectively.

Table 6 Model properties of the six-story shear-type building, synthetic building data

| Mode | 1 | 2 | 3 | 4 | 5 | 6 |
|---|---|---|---|---|---|---|
| Frequency (Hz) | 2.87 | 2.96 | 3.21 | 8.44 | 8.71 | 9.45 |
| Damping ratio (%) | 0.5 | 0.5 | 1.0 | 1.0 | 1.2 | 1.2 |
| MPF | 0.0096 | 0.0080 | 0.0080 | 0.0090 | 0.0075 | 0.0075 |
| Young's modulus (GPa) | 70 | | | | | |
| Poisson's ratio | 0.3 | | | | | |
| Mass density (g/mm$^3$) | 2.7 | | | | | |
| Column's section area (mm$^3$) | 16.5 | | | | | |
| Column's moment of inertia (mm$^4$) | $I_x = 141.667; I_x = 166.375; J = 17$ | | | | | |

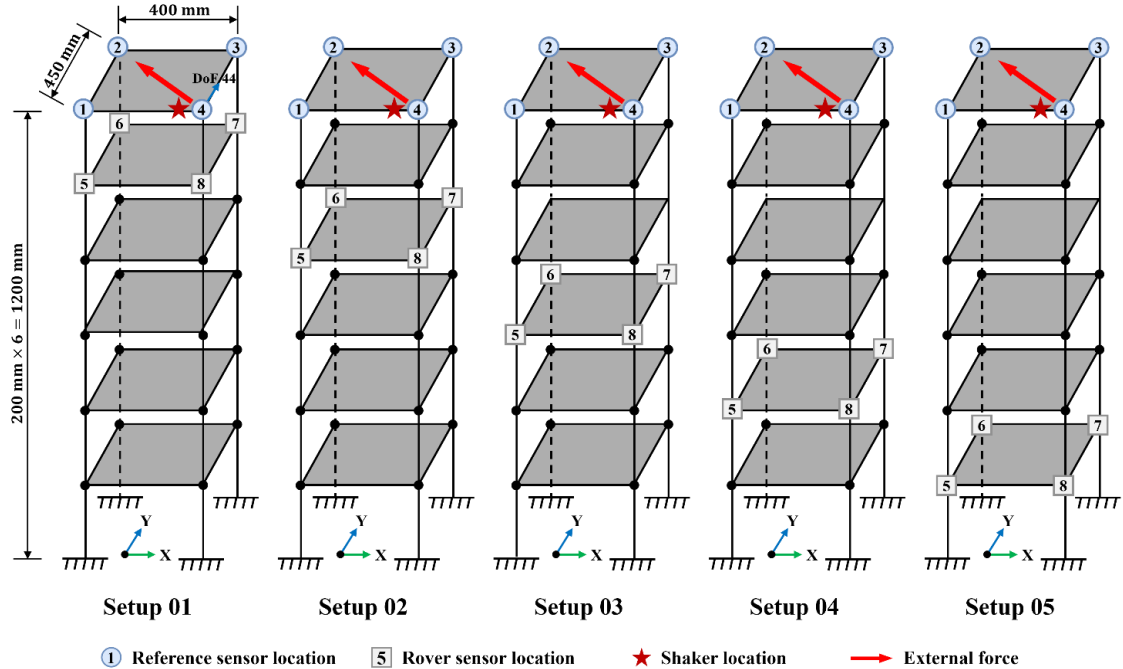

Fig. 5 Geometric property of the six-story shear-type building and test configuration, synthetic building data

The proposed MS-BAYEMA algorithm is adopted for identification, with identified results shown in Fig. 7. The identified modal parameters closely resemble the true values, demonstrating the validity of MS-BAYEMA. Notably, while identifying the damping ratio is generally more challenging than the natural frequency, the identified damping ratios are considerably reliable, indicated by the small c.o.v.s (around 1%). As the main purpose of the multi-setup vibration test, the global mode shape can be accurately identified, as evidenced by the high scores of MAC and small values of identification

uncertainty. In addition, the MPFs are identified to be 0.0095, 0.0080, 0.0080, 0.0090, 0.0071, 0.0072, respectively, which are all close to the true values.

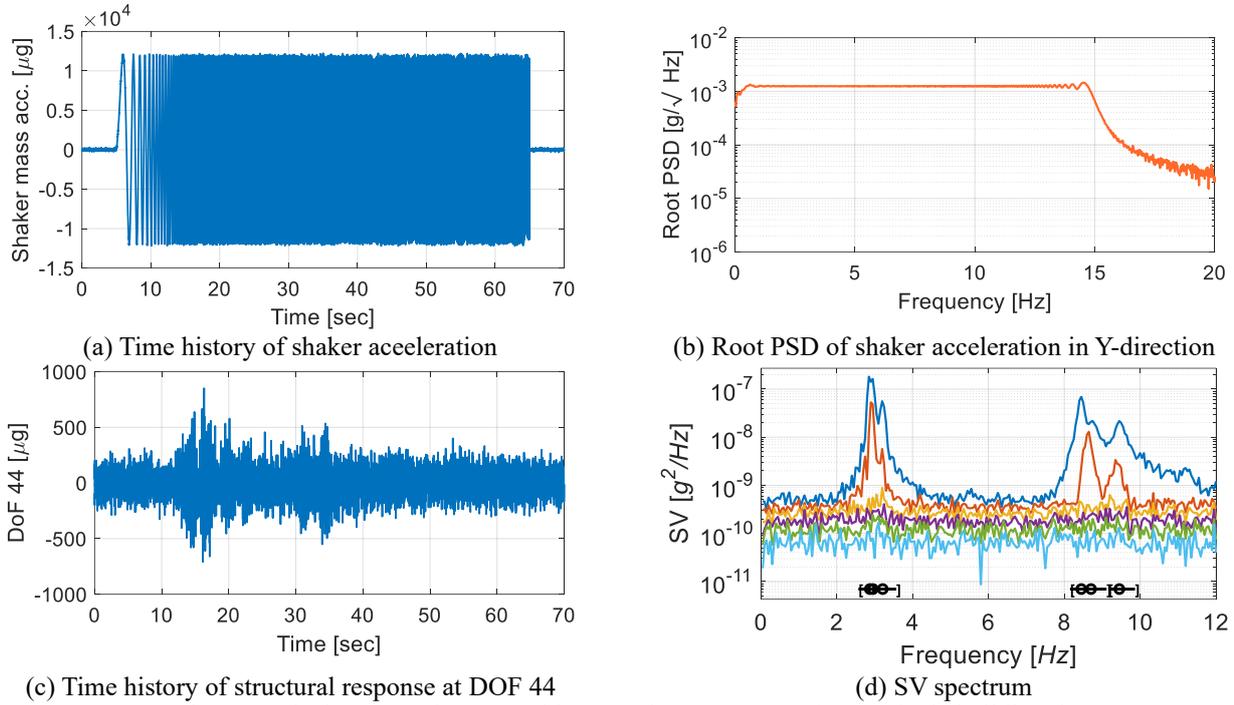

(a) Time history of shaker acceleration  
(b) Root PSD of shaker acceleration in Y-direction  
(c) Time history of structural response at DoF 44  
(d) SV spectrum  
Fig. 6 Typical measured structural input and output, Setup 01, synthetic building data

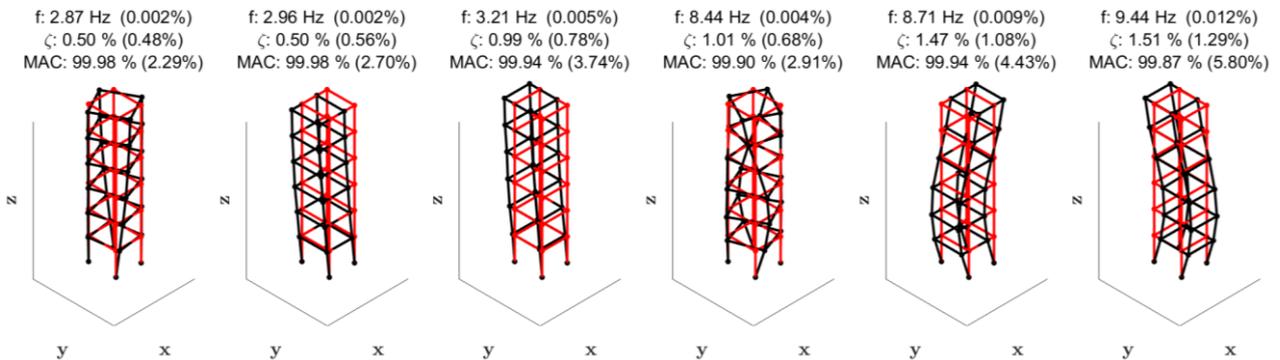

Fig. 7 Identified results for 6 lateral modes; six-story shear-type building, synthetic building data

In the development of the MS-BAYEMA algorithm, the shaker acceleration is assumed to be measured without error. To validate this assumption, a parametric study is conducted by introducing varying levels of measurement noise. More specifically, four levels of white noise (0.1, 1, 10 and 100 $\mu g/\sqrt{Hz}$) are superimposed to the raw input. They correspond to a very good servo-accelerometer (e.g., professional seismograph), a good servo-accelerometer, a good piezoelectric accelerometer and a poor accelerometer (e.g., used in smart phones), respectively. The normalized frequencies and damping ratios (identified values / nominal values) as well as MAC are shown in Fig. 8. In the figure, the dashed lines represent the normalized values equal to 1, although they are translated for display. The well-identified results (in terms of MAPs and c.o.v.s) indicate that neglecting the measurement noise in the input has almost no influence on the results, provided that the external acceleration significantly exceeds the sensor noise level. However, if low-quality accelerometers are used, it may introduce bias into the results. Nevertheless, this issue can be effectively addressed by either increasing the input energy or employing higher-quality sensors. Incorporating measurement noise directly into the input modeling may unnecessarily complicate the problem.

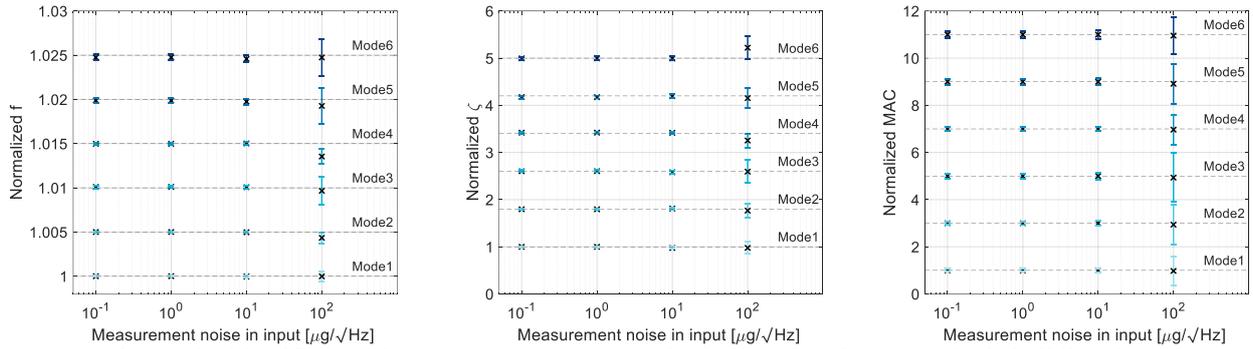

Fig. 8 The identified results versus measurement noise in measured input, synthetic building data

## 4.2 Application with field data

This section consists of two field test examples, based on a steel-truss footbridge and an engineered bamboo building. They are used to illustrate the practicality of the proposed algorithm under operational conditions.

### 4.2.1 A steel-truss footbridge

The instrumented structure is a steel-truss footbridge located in the International Campus of Zhejiang University, as shown in Fig 9(a). The test configuration details are shown in Fig 9(d), measuring a total of 14 DOFs of the structure along the Z-direction. A broadband pseudo-random force with a flat root acceleration PSD was generated using an electromagnetic shaker with a moving mass of 30.6 kg, as depicted in Fig. 9(b). The acceleration excitation and structural responses were recorded by 15 piezoelectric accelerometers (Fig 9(c)) originally sampled at 2048 Hz but later decimated to 128 Hz for processing. Further details about the experiment and the corresponding data can be found in Ref. [16]. To fully evaluate the algorithm's performance, we consider four cases for comparison, as listed in Table 7. It is important to note that the original field tests consist of two single-setup tests, specifically Case 02 and Case 04. The multi-setup vibration data, represented by Case 01 and Case 03, are artificially generated from these two original datasets.

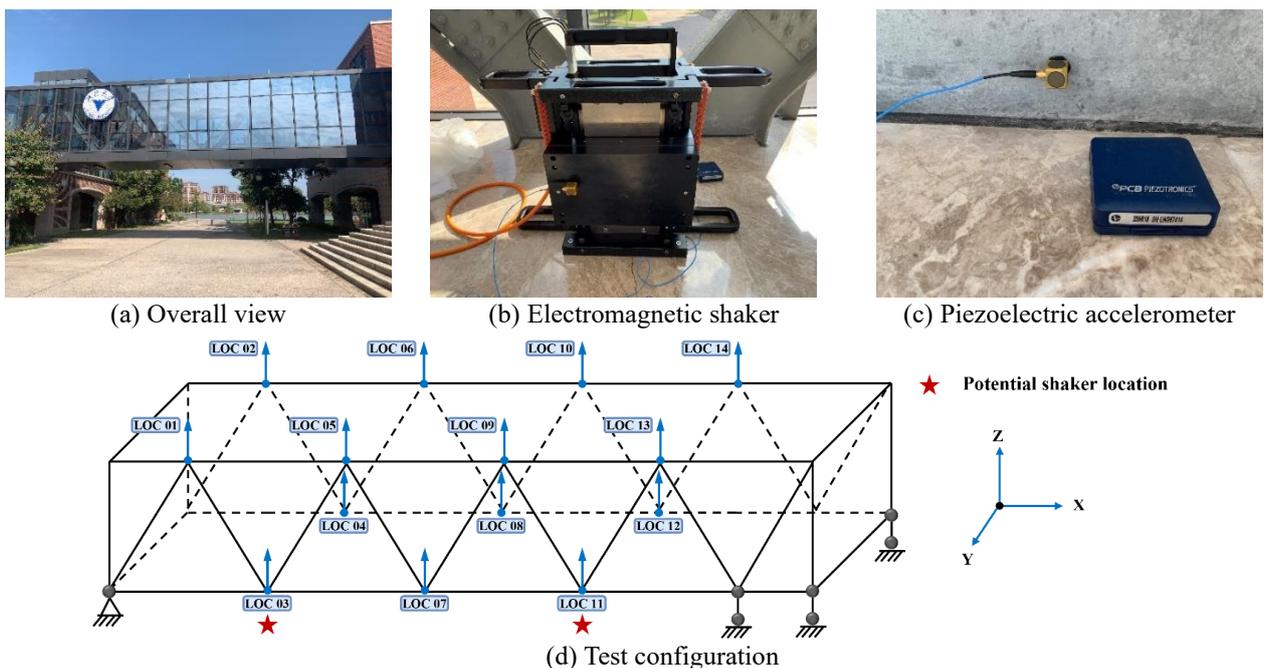

Fig. 9 A steel-truss footbridge in the International Campus, Zhejiang University, field bridge data

Table 7 Test configuration for different cases, field bridge data

| Case No. | Setup No. | Rover DOF | Test duration (sec) | Shaker DOF | Reference DOF | Used method |
|---|---|---|---|---|---|---|
| 01 | 1 | 1,2 | 60 | 11 | 9,10 | MS-BAYEMA |
|    | 2 | 3,4 | 60 | | | |
|    | 3 | 5,6 | 60 | | | |
|    | 4 | 7,8 | 60 | | | |
|    | 5 | 11,12 | 60 | | | |
|    | 6 | 13,14 | 60 | | | |
| 02 | 1 | All | 360 | 11 | All | SS-BAYEMA |
| 03 | 1 | 1,2 | 60 | 3 | 9,10 | MS-BAYEMA |
|    | 2 | 3,4 | 60 | | | |
|    | 3 | 5,6 | 60 | | | |
|    | 4 | 7,8 | 60 | | | |
|    | 5 | 11,12 | 60 | | | |
|    | 6 | 13,14 | 60 | | | |
| 04 | 1 | All | 360 | 3 | All | SS-BAYEMA |

We first verify the proposed algorithm using data of Case 01. The root PSD spectrum of the input and the root SV spectrum of the structural responses for Case 01 are presented in Fig. 10, where clear picks indicate three well-excited modes within [0, 20] Hz. These modes are selected as three well-separated modes for modal identification using MS-BAYEMA, with brackets marking the frequency bands and circles indicating the initial guesses of natural frequencies. The first three vertical modes identified by MS-BAYEMA are shown in Fig. 11. It is seen that these three modes correspond to the first vertical symmetric mode, first torsional mode and first vertical asymmetric mode, respectively. The identified MPFs of three modes are 0.0011, 0.0009, and 0.0010, respectively. These results demonstrate that a small shaker with a moving mass of 30.6 kg can effectively excite a bridge with a modal mass of nearly 10 tons to an acceptable vibration magnitude, which highlights the feasibility of multiple-setup EMA for the moderate-sized structures.

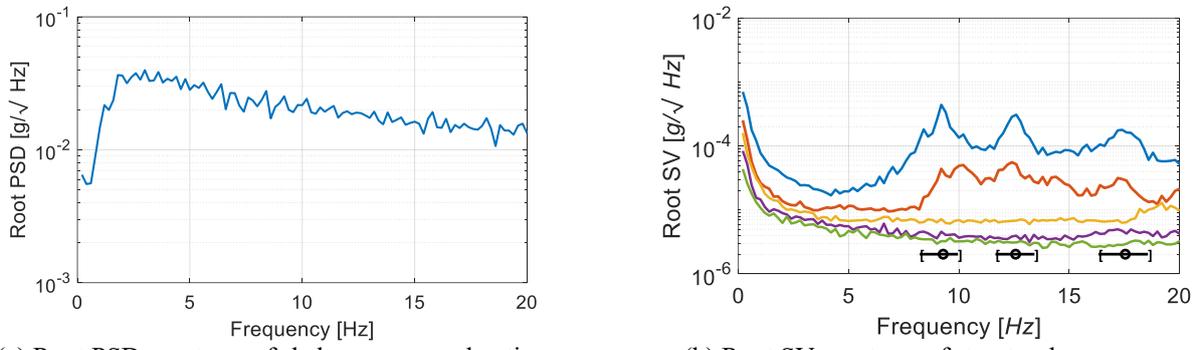

(a) Root PSD spectrum of shaker mass acceleration   (b) Root SV spectrum of structural responses

Fig. 10  Root PSD of measured input and root SV of measured output, Setup 01 in Case 01, field bridge data

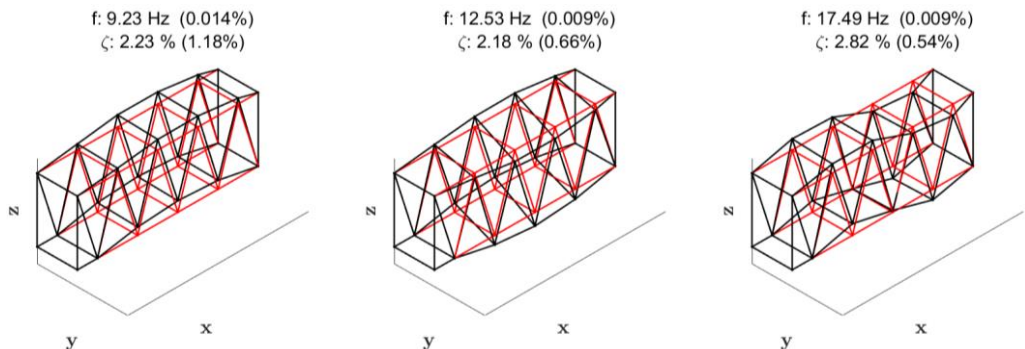

Fig. 11  Identified results for first three vertical modes, field bridge data

In order to further investigate the algorithm's performance, we apply the proposed algorithm to all other cases. The selected frequency bands and the initial guesses for natural frequencies and damping ratios are maintained the same in all cases. consistently. The identified results are summarized in Table. 8. Note that the MAC is calculated relative to identified mode shapes by SS-BAYEMA in Case 02. For Case 01 and Case 02, identified modal parameters are closely resemble each other, despite minor difference in damping ratio. As the main concern of MS-BAYEMA, three MAC values of 99.99% indicate its ability to successfully identify all global mode shapes. In terms of the identification uncertainty, SS-BAYEMA yields smaller values for all identified modal parameters, attributed to the larger number of sensors used. However, the slight increase in identification uncertainty with MS-BAYEMA is negligible in engineering applications, especially when considering the significant costs in sensor instrumentation required for SS-BAYEMA. For Case 03 and Case 04, similar performance can be observed. They both illustrate that the proposed MS-BAYEMA algorithm is highly effective in overcoming challenges associated with limited sensor instrumentation.

Table 8 identified modal parameters for all cases, field bridge data

| Parameter | Natural frequency (Hz) | | | Damping ratio (%) | | | Mode shape MAC (%) | | |
|---|---|---|---|---|---|---|---|---|---|
| Mode | 1 | 2 | 3 | 1 | 2 | 3 | 1 | 2 | 3 |
| Case 01 | 9.23 | 12.53 | 17.49 | 2.23 | 2.18 | 2.82 | 99.99 | 99.99 | 99.99 |
| (MS-BAYEMA) | (0.014%) | (0.009%) | (0.009%) | (1.18%) | (0.66%) | (0.54%) | (2.26%) | (1.60%) | (1.22%) |
| Case 02 | 9.23 | 12.53 | 17.50 | 2.23 | 2.16 | 2.80 | 100.00 | 100.00 | 100.00 |
| (SS-BAYEMA) | (0.007%) | (0.005%) | (0.005%) | (0.61%) | (0.40%) | (0.31%) | (0.88%) | (0.70%) | (0.53%) |
| Case 03 | 9.21 | 12.55 | 17.48 | 2.92 | 2.17 | 2.76 | 99.99 | 99.98 | 99.99 |
| (MS-BAYEMA) | (0.022%) | (0.011%) | (0.010%) | (1.59%) | (0.77%) | (0.61%) | (2.92%) | (2.44%) | (1.57%) |
| Case 04 | 9.20 | 12.56 | 17.47 | 3.39 | 2.18 | 2.80 | 99.99 | 100.00 | 99.99 |
| (SS-BAYEMA) | (0.014%) | (0.008%) | (0.006%) | (0.82%) | (0.60%) | (0.39%) | (1.14%) | (1.05%) | (0.64%) |

Note: MAC is calculated based on the identified mode shapes by SS-BAYEMA in Case 02.

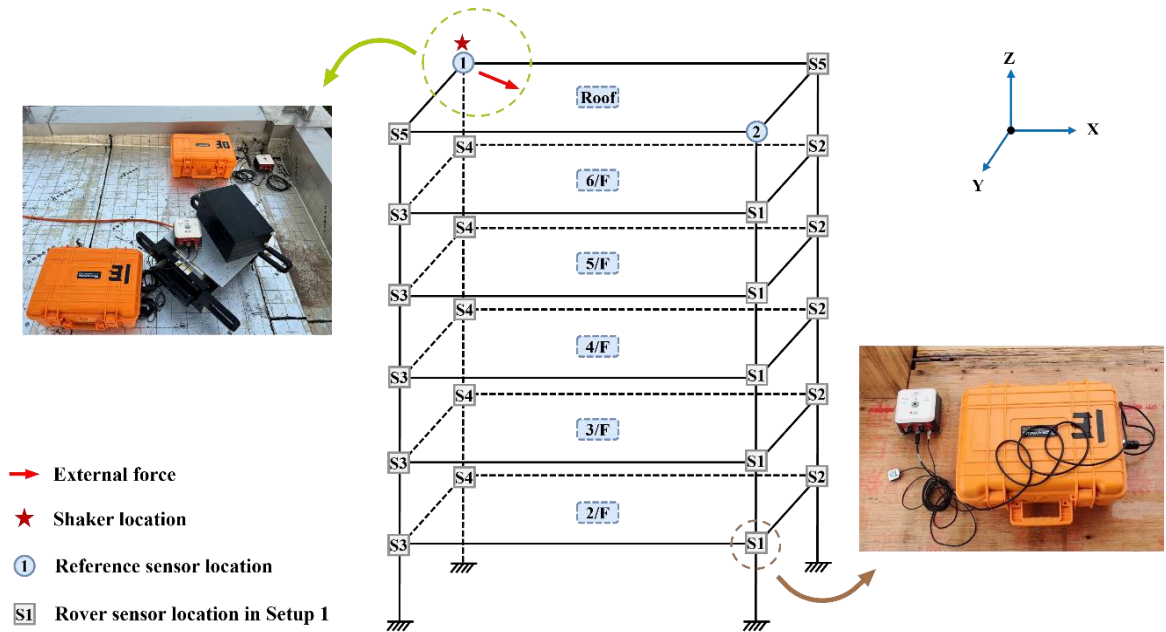

Fig. 12 Geometrical properties of the instrumented building structure and test configuration, field building data

### 4.2.2 A six-story engineered bamboo building

The final examined example is a multi-setup EMA of a six-story engineered bamboo building. The detailed description of the building can be found in Ref. [40]. The test configuration is illustrated in Fig 12. The electromagnetic shaker was placed diagonally on the roof floor to generate excitations in both X- and Y-directions. A broadband pseudo-random force with a flat root acceleration PSD was generated by the shaker with a moving mass of 32.1 kg. Seven triaxial servo-accelerometers (Kinemetrics ETNA2) were used recorded the structural dynamic responses in X- and Y-directions at four corners of each of the six floors, resulting in a total of 48 DOFs to be measured. To be more specific, five of these accelerometers were movable and served as rover sensors across different setups, while the remaining two were fixed as reference sensors. Accordingly, a five-setup forced vibration test was conducted on the structure, with the setup number "S1"-"S5" indicated in Fig 12.

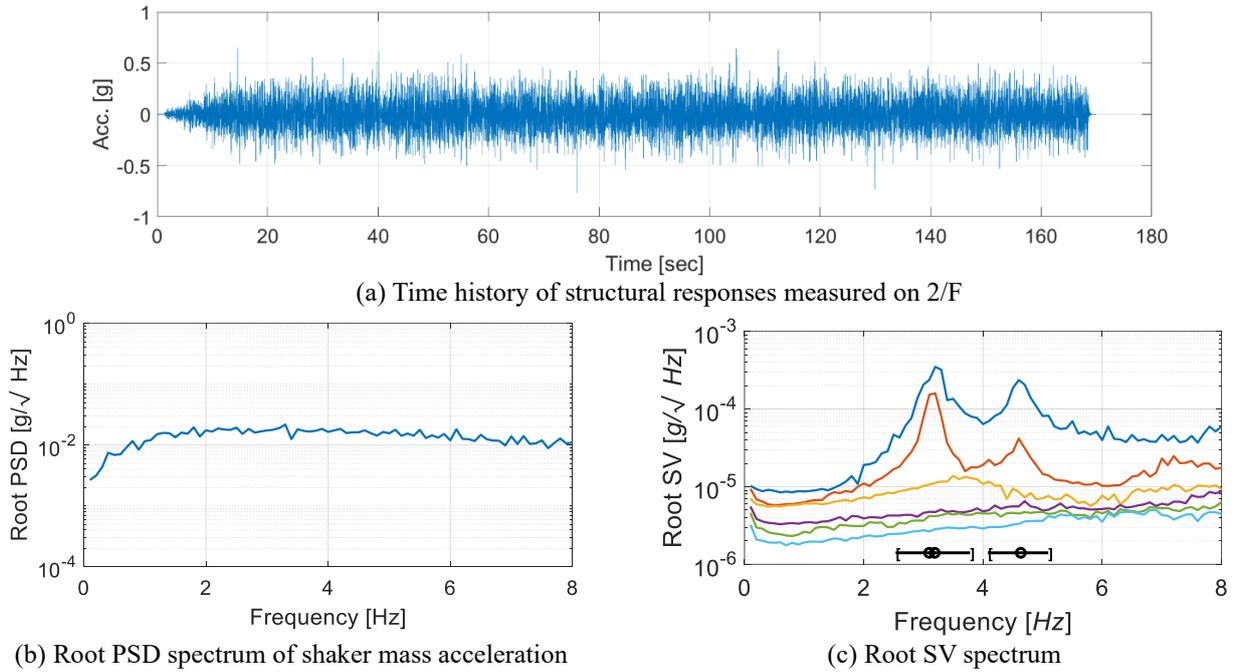

(a) Time history of structural responses measured on 2/F

(b) Root PSD spectrum of shaker mass acceleration  (c) Root SV spectrum

Fig. 13 Typical structural input and output; Setup 01, field building data

Using the structural input and output vibration data in the selected frequency bands, MS-BAYEMA is applied to identify these three modes. The identified results are illustrated in Fig. 14, including two translational modes along X- and Y- directions, as well as a torsional mode. It is worth mentioning that the first two modes are well-identified, despite being closely-spaced modes. This demonstrates the proposed approach's capability to handle such a challenging task in modal analysis. Moreover, the identified MPFs are calculated to be 0.0014, 0.0015, and 0.0012, respectively. It is seen that a relatively small shaker with a moving mass of 32.1 kg can effectively excite a building with a modal mass of around 10 tons to a sufficient level for modal identification.

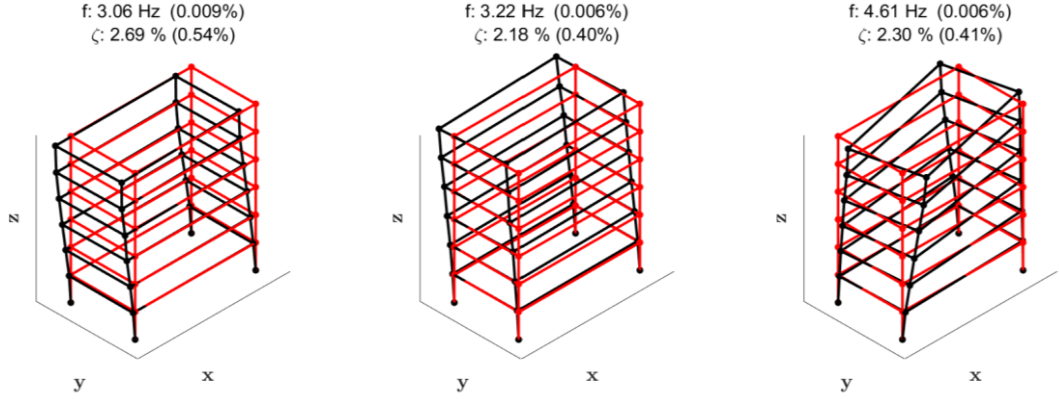

Fig. 14 Identified results for first three translational modes, field building data

## 5 Conclusions

Aiming at breaking down the limitations on the quantity of sensors and excitation requirement in EMA, this paper develops a Bayesian FFT modal identification algorithm using multi-setup forced vibration data. By virtue of complex matrix calculus, the MAP and PCM of modal parameters are derived analytically and calculated with high efficiency. By investigating representative examples with synthetic and field data, the validation and applicability of the algorithm are thoroughly illustrated. More specifically, several key conclusions are summarized as follows:

(1) The proposed MS-BAYEMA algorithm can obtain high-spatial-resolution mode shape using a limited number of sensors. It makes EMA insensitive to the quantity of the test equipment and therefore more widely applicable.
(2) Compared to existing SS-BAYEMA, the proposed MS-BAYEMA can obtain more accurate identification when dealing with the structural modes that closed or coupled in multiple directions. This improvement is attributed to MS-BAYEMA's capability to incorporate input and output vibration data from different directions even when a single-axial shaker is used.
(3) The field examples show that a small shaker (around 30 kg) can effectively excite structures with a modal mass of 10 tons, demonstrating the feasibility of multi-setup EMA.

With the identified high-resolution mode shapes, a refined input-output model of the tested structure is effectively constructed. It offers a robust framework for predicting the structural response under various future loading conditions, thereby enabling a more precise assessment of the structure's current condition and its potential behavior over time.

## ACKNOWLEDGMENTS

This research is supported by the National Natural Science Foundation of China (U23A20662).

## Appendix A  Proof of gradient of NLLF

This appendix analytically derives the condensed form of the gradient of $\mathcal{L}(\boldsymbol{\theta})$ w.r.t. $\boldsymbol{\Phi}$, $\boldsymbol{\Lambda}^{(s)}$ and $S_e^{(r)}$. The key results in matrix theory that are essential for the derivation can be found in Appendix A of Ref. [13]. Firstly, we recall that the negative log-likelihood function $\mathcal{L}(\boldsymbol{\theta})$ in Eqn. (8) is given by

$$\mathcal{L}(\boldsymbol{\theta}) = \sum_{r=1}^{n_r} d_r N_f^{(r)} \ln \pi + \sum_{r=1}^{n_r} d_r N_f^{(r)} \ln S_e^{(r)} + \sum_{r=1}^{n_r} S_e^{(r)^{-1}} \sum_{k=1}^{N_f^{(r)}} A_k^{(r)}(\boldsymbol{\theta}) \qquad (A.1)$$

where

$$A_k^{(r)}(\boldsymbol{\theta}) = [\widehat{\boldsymbol{Y}}_k^{(r)} - \boldsymbol{S}^{(r)}\boldsymbol{\Phi} \boldsymbol{H}_k^{(r)} \boldsymbol{\Lambda}^{(S(r))\mathrm{T}} \ddot{\boldsymbol{U}}_k^{(r)}]^* [\widehat{\boldsymbol{Y}}_k^{(r)} - \boldsymbol{S}^{(r)}\boldsymbol{\Phi} \boldsymbol{H}_k^{(r)} \boldsymbol{\Lambda}^{(S(r))\mathrm{T}} \ddot{\boldsymbol{U}}_k^{(r)}] \quad (A.2)$$

We take the first-order derivative of the NLLF w.r.t. different parameters and setting them to zero so that we can obtain their update. The detailed derivation is as follows.

1. Error term PSD $S_e^{(r)}$

To obtain the update of $S_e^{(r)}$, taking the first-order partial derivative of $\mathcal{L}(\boldsymbol{\theta})$ w.r.t. $S_e^{(r)}$ gives

$$\frac{\partial \mathcal{L}(\boldsymbol{\theta})}{\partial S_e^{(r)}} = \frac{d_r N_f^{(r)}}{S_e^{(r)}} - \frac{\sum_{k=1}^{N_f^{(r)}} A_k^{(r)}(\boldsymbol{\theta})}{S_e^{(r)2}} \quad (A.3)$$

Setting it to zero and solving for $S_e^{(r)}$ yields

$$S_e^{(r)} = \frac{\sum_{k=1}^{N_f^{(r)}} A_k^{(r)}(\boldsymbol{\theta})}{d_r N_f^{(r)}} \quad (A.4)$$

2. Mode shape matrix $\boldsymbol{\Phi}$

Re-write Eqn. (A.2) as

$$A_k^{(r)}(\boldsymbol{\theta}) = \mathrm{tr}[\widehat{\boldsymbol{Y}}_k^{(r)} \widehat{\boldsymbol{Y}}_k^{(r)*}] - \mathrm{tr}[2\mathrm{Re}(\boldsymbol{S}^{(r)} \boldsymbol{\Phi} \boldsymbol{H}_k^{(r)} \boldsymbol{\Lambda}^{(S(r))\mathrm{T}} \ddot{\boldsymbol{U}}_k^{(r)} \widehat{\boldsymbol{Y}}_k^{(r)*})] + \mathrm{tr}[\boldsymbol{S}^{(r)} \boldsymbol{\Phi} \boldsymbol{B}_k^{(r)} \boldsymbol{\Phi}^\mathrm{T} \boldsymbol{S}^{(r)\mathrm{T}}] \quad (A.5)$$

where

$$\boldsymbol{B}_k^{(r)} = \boldsymbol{H}_k^{(r)} \boldsymbol{\Lambda}^{(S(r))\mathrm{T}} \ddot{\boldsymbol{U}}_k^{(r)} \ddot{\boldsymbol{U}}_k^{(r)*} \boldsymbol{\Lambda}^{(S(r))} \boldsymbol{H}_k^{(r)*} \quad (A.6)$$

To obtain the update of $\boldsymbol{\Phi}$, taking the first-order partial derivative of $\mathcal{L}(\boldsymbol{\theta})$ w.r.t. $\boldsymbol{\Phi}$ gives

$$\frac{\partial \mathcal{L}(\boldsymbol{\theta})}{\partial \boldsymbol{\Phi}} = \sum_{r=1}^{n_r} S_e^{(r)-1} \left\{ -2\mathrm{Re}\left[ \mathrm{vec}^\mathrm{T} \left( \sum_{k=1}^{N_f^{(r)}} \boldsymbol{S}^{(r)\mathrm{T}} \widehat{\boldsymbol{Y}}_k^{(r)} \ddot{\boldsymbol{U}}_k^{(r)*} \boldsymbol{\Lambda}^{(S(r))} \boldsymbol{H}_k^{(r)*} \right) \right] + 2\mathrm{vec}^\mathrm{T}(\boldsymbol{\Phi}) \left[ \mathrm{Re}\left( \sum_{k=1}^{N_f^{(r)}} \boldsymbol{B}_k^{(r)} \right) \otimes (\boldsymbol{S}^{(r)\mathrm{T}} \boldsymbol{S}^{(r)}) \right] \right\} \quad (A.7)$$

where "vec($\boldsymbol{\Phi}$)" denotes a vector by stacking the columns of $\boldsymbol{\Phi}$. Setting it to zero and solving for $\boldsymbol{\Phi}$ yields

$$\mathrm{vec}^\mathrm{T}(\boldsymbol{\Phi}) = \frac{\sum_{r=1}^{n_r} S_e^{(r)-1} \mathrm{Re}\left[ \mathrm{vec}^\mathrm{T} \left( \sum_{k=1}^{N_f^{(r)}} \boldsymbol{S}^{(r)\mathrm{T}} \widehat{\boldsymbol{Y}}_k^{(r)} \ddot{\boldsymbol{U}}_k^{(r)*} \boldsymbol{\Lambda}^{(S(r))} \boldsymbol{H}_k^{(r)*} \right) \right]}{\sum_{r=1}^{n_r} S_e^{(r)-1} \left[ \mathrm{Re}\left( \sum_{k=1}^{N_f^{(r)}} \boldsymbol{B}_k^{(r)} \right) \otimes (\boldsymbol{S}^{(r)\mathrm{T}} \boldsymbol{S}^{(r)}) \right]} \quad (A.8)$$

3. Modal participation factor $\boldsymbol{\Lambda}^{(s)}$

Re-write Eqn. (A.2) as

$$A_k^{(r)}(\boldsymbol{\theta}) = \mathrm{tr}[\widehat{\boldsymbol{Y}}_k^{(r)} \widehat{\boldsymbol{Y}}_k^{(r)*}] - \mathrm{tr}[2\mathrm{Re}(\widehat{\boldsymbol{Y}}_k^{(r)} \ddot{\boldsymbol{U}}_k^{(r)*} \boldsymbol{\Lambda}^{(S(r))} \boldsymbol{H}_k^* \boldsymbol{\Phi}^\mathrm{T} \boldsymbol{S}^{(r)\mathrm{T}})] + \mathrm{tr}[\ddot{\boldsymbol{U}}_k^{(r)*} \boldsymbol{\Lambda}^{(S(r))} \boldsymbol{C}_k^{(r)} \boldsymbol{\Lambda}^{(S(r))\mathrm{T}} \ddot{\boldsymbol{U}}_k^{(r)}] \quad (A.9)$$

where

$$\boldsymbol{C}_k^{(r)} = \boldsymbol{H}_k^{(r)*} \boldsymbol{\Phi}^\mathrm{T} \boldsymbol{S}^{(r)\mathrm{T}} \boldsymbol{S}^{(r)} \boldsymbol{\Phi} \boldsymbol{H}_k^{(r)} \quad (A.10)$$

To obtain the updated value of $\boldsymbol{\Lambda}^{(s)}$, taking the first-order partial derivatives of $\mathcal{L}(\boldsymbol{\theta})$ w.r.t. $\boldsymbol{\Lambda}^{(s)}$ gives

$$\frac{\partial \mathcal{L}(\boldsymbol{\theta})}{\partial \boldsymbol{\Lambda}^{(s)}} = \sum_{r \in \mathcal{R}(s)} S_e^{(r)-1} \left[ -2\mathrm{vec}^\mathrm{T}\left( \mathrm{Re} \sum_k \ddot{\boldsymbol{U}}_k^{(r)} \widehat{\boldsymbol{Y}}_k^{(r)*} \boldsymbol{S}^{(r)} \boldsymbol{\Phi} \boldsymbol{H}_k^{(r)} \right) \right] + \sum_{r \in \mathcal{R}(s)} S_e^{(r)-1} 2\mathrm{vec}^\mathrm{T}(\boldsymbol{\Lambda}^{(s)}) \mathrm{Re} \sum_{k=1}^{N_f^{(r)}} [\boldsymbol{C}_k^{(r)\mathrm{T}} \otimes (\ddot{\boldsymbol{U}}_k^{(r)} \ddot{\boldsymbol{U}}_k^{(r)*})] \quad (A.11)$$

Setting it to zero and solving for $\boldsymbol{\Lambda}^{(s)}$ yields

$$\mathrm{vec}(\boldsymbol{\Lambda}^{(s)}) = \left\{ \sum_{r \in \mathcal{R}(s)} S_e^{(r)-1} \mathrm{Re} \sum_{k=1}^{N_f^{(r)}} [\boldsymbol{C}_k^{(r)\mathrm{T}} \otimes (\ddot{\boldsymbol{U}}_k^{(r)} \ddot{\boldsymbol{U}}_k^{(r)*})] \right\}^{-1} \left\{ \sum_{r \in \mathcal{R}(s)} S_e^{(r)-1} \mathrm{Re} \sum_{k=1}^{N_f^{(r)}} [\boldsymbol{H}_k^{(r)} \otimes (\ddot{\boldsymbol{U}}_k^{(r)} \widehat{\boldsymbol{Y}}_k^{(r)*})] \mathrm{vec}(\boldsymbol{S}^{(r)} \boldsymbol{\Phi}) \right\} \quad (A.12)$$

# Appendix B  Proof of Hessian matrix of NLLF

This appendix derives the Hessian matrix of $\mathcal{L}(\boldsymbol{\theta})$ w.r.t. $\boldsymbol{\Phi}$, $\boldsymbol{\Lambda}^{(s)}$ and $S_e^{(r)}$ and $\boldsymbol{\theta}_m = [f_1, \zeta_1, \ldots, f_m, \zeta_m]^T$. The used key results in linear algebra and complex matrix calculus can be found in Appendix A of Ref. [13]. In addition, the first-order and second-order derivatives of frequency response function are listed in Appendix C.

1. Diagonal terms

We first derive the diagonal terms of the Hessian matrix $\nabla^2 \mathcal{L}(\boldsymbol{\theta})$ following basic rules of complex matrix calculus.

(1) Following Eqn. (A.3), the second-order derivative of NLLF w.r.t. $S_e^{(r)}$ gives

$$\frac{\partial^2 \mathcal{L}(\boldsymbol{\theta})}{\partial S_e^{(r)^2}} = -d_r N_f^{(r)} S_e^{(r)-2} + 2 S_e^{(r)-3} \sum_{k=1}^{N_f^{(r)}} A_k^{(r)}(\boldsymbol{\theta}) \tag{B.1}$$

(2) Following Eqn. (A.7), taking the second-order derivative of NLLF w.r.t. $\boldsymbol{\Phi}$ yields

$$\frac{\partial^2 \mathcal{L}(\boldsymbol{\theta})}{\partial \boldsymbol{\Phi} \partial \boldsymbol{\Phi}^T} = \frac{\partial}{\partial \text{vec}(\boldsymbol{\Phi})} \sum_{r=1}^{n_r} 2 S_e^{(r)-1} \left\{ \left[ \text{Re}\left( \sum_{k=1}^{N_f^{(r)}} \boldsymbol{B}_k^{(r)} \right) \otimes \left( \boldsymbol{S}^{(r)T} \boldsymbol{S}^{(r)} \right) \right] \text{vec}(\boldsymbol{\Phi}) \right\}$$

$$= \sum_{r=1}^{n_r} 2 S_e^{(r)-1} \left[ \text{Re}\left( \sum_{k=1}^{N_f^{(r)}} \boldsymbol{B}_k^{(r)} \right) \otimes \left( \boldsymbol{S}^{(r)T} \boldsymbol{S}^{(r)} \right) \right] \tag{B.2}$$

(3) Following Eqn. (A.11), taking the second-order derivative of NLLF w.r.t. $\boldsymbol{\Lambda}^{(s)}$ gives

$$\frac{\partial^2 \mathcal{L}(\boldsymbol{\theta})}{\partial \boldsymbol{\Lambda}^{(s)} \partial \boldsymbol{\Lambda}^{(s)T}} = \frac{\partial}{\partial \text{vec}(\boldsymbol{\Lambda}^{(s)})} \sum_{r \in \mathcal{R}(s)} S_e^{(r)-1} \left( 2 \left[ \text{Re}\left( \sum_{k=1}^{N_f^{(r)}} \boldsymbol{C}_k^{(r)} \right) \otimes \left( \overline{\ddot{\boldsymbol{U}}_k^{(r)} \ddot{\boldsymbol{U}}_k^{(r)*}} \right) \right] \text{vec}(\boldsymbol{\Lambda}^{(s)}) \right)$$

$$= \sum_{r \in \mathcal{R}(s)} 2 S_e^{(r)-1} \left[ \text{Re}\left( \sum_{k=1}^{N_f^{(r)}} \boldsymbol{C}_k^{(r)T} \right) \otimes \left( \ddot{\boldsymbol{U}}_k^{(r)} \ddot{\boldsymbol{U}}_k^{(r)*} \right) \right] \tag{B.3}$$

(4) The second-order derivative of NLLF w.r.t. $\boldsymbol{\theta}_m$

For $\boldsymbol{\theta}_m$ encoded in the FRF $\boldsymbol{H}_k^{(r)}$, we employ the chain rule and treat $\boldsymbol{H}_k^{(r)}$ and its conjugate $\overline{\boldsymbol{H}}_k^{(r)}$ as independent variables in the following analysis. The first-order derivative then reads

$$\frac{\partial \mathcal{L}(\boldsymbol{\theta})}{\partial \boldsymbol{\theta}_m} = \sum_{r=1}^{n_r} \sum_{k=1}^{N_f^{(r)}} \left[ \frac{\partial \mathcal{L}(\boldsymbol{\theta})}{\partial \text{vec}(\boldsymbol{H}_k^{(r)})} \quad \frac{\partial \mathcal{L}(\boldsymbol{\theta})}{\partial \text{vec}(\overline{\boldsymbol{H}}_k^{(r)})} \right] \begin{bmatrix} \frac{\partial \text{vec}(\boldsymbol{H}_k^{(r)})}{\partial \boldsymbol{\theta}_m} \\ \frac{\partial \text{vec}(\overline{\boldsymbol{H}}_k^{(r)})}{\partial \boldsymbol{\theta}_m} \end{bmatrix} \tag{B.4}$$

Note that the FRF $\boldsymbol{H}_k^{(r)}$ is a diagonal matrix, it is possible to find a commutation matrix $\boldsymbol{L}_d \in \mathbb{R}^{m^2 \times m}$ consisting solely of zeros and ones such that

$$\begin{cases} \text{vec}(\boldsymbol{H}_k^{(r)}) = \boldsymbol{L}_d \text{vec}_d(\boldsymbol{H}_k^{(r)}) \\ \text{vec}_d(\boldsymbol{H}_k^{(r)}) = \boldsymbol{L}_d^T \text{vec}(\boldsymbol{H}_k^{(r)}) \\ \boldsymbol{L}_d^T \boldsymbol{L}_d = \boldsymbol{I}_m \end{cases} \tag{B.5}$$

where "$\text{vec}_d(\boldsymbol{H}_k^{(r)})$" denotes a vector by stacking the diagonal elements of $\boldsymbol{H}_k^{(r)}$. Therefore,

$$\frac{\partial \mathcal{L}(\boldsymbol{\theta})}{\partial \boldsymbol{\theta}_m} = \sum_{r=1}^{n_r} \sum_{k=1}^{N_f^{(r)}} \left[ \frac{\partial \mathcal{L}(\boldsymbol{\theta})}{\partial \text{vec}(\boldsymbol{H}_k^{(r)})} \quad \frac{\partial \mathcal{L}(\boldsymbol{\theta})}{\partial \text{vec}(\overline{\boldsymbol{H}}_k^{(r)})} \right] \cdot (\boldsymbol{I}_2 \otimes \boldsymbol{L}_d) \cdot \begin{bmatrix} \frac{\partial \text{vec}_d(\boldsymbol{H}_k^{(r)})}{\partial \boldsymbol{\theta}_m} \\ \frac{\partial \text{vec}_d(\overline{\boldsymbol{H}}_k^{(r)})}{\partial \boldsymbol{\theta}_m} \end{bmatrix} \tag{B.6}$$

where

$$\begin{cases} \frac{\partial \mathcal{L}(\boldsymbol{\theta})}{\partial \text{vec}(\boldsymbol{H}_k^{(r)})} = S_e^{(r)-1} \text{vec}^{\text{T}} \left[ -\boldsymbol{\Phi}^{\text{T}} \boldsymbol{S}^{(r)\text{T}} \left( \widehat{\overline{\boldsymbol{Y}}}_k^{(r)} \ddot{\boldsymbol{U}}_k^{(r)*} - \boldsymbol{S}^{(r)} \boldsymbol{\Phi} \overline{\boldsymbol{H}}_k^{(r)} \boldsymbol{\Lambda}^{(S(r))} \overline{\ddot{\boldsymbol{U}}}_k^{(r)} \ddot{\boldsymbol{U}}_k^{(r)*} \right) \boldsymbol{\Lambda}^{(S(r))} \right] \\ \frac{\partial \mathcal{L}(\boldsymbol{\theta})}{\partial \text{vec}(\overline{\boldsymbol{H}}_k^{(r)})} = S_e^{(r)-1} \text{vec}^{\text{T}} \left[ -\boldsymbol{\Phi}^{\text{T}} \boldsymbol{S}^{(r)\text{T}} (\widehat{\boldsymbol{Y}}_k^{(r)} \ddot{\boldsymbol{U}}_k^{(r)*} - \boldsymbol{S}^{(r)} \boldsymbol{\Phi} \boldsymbol{H}_k^{(r)} \boldsymbol{\Lambda}^{(S(r))} \ddot{\boldsymbol{U}}_k^{(r)} \ddot{\boldsymbol{U}}_k^{(r)*}) \boldsymbol{\Lambda}^{(S(r))} \right] \end{cases} \tag{B.7}$$

and

$$\frac{\partial \text{vec}_d(\boldsymbol{H}_k^{(r)})}{\partial \boldsymbol{\theta}_m^{\text{T}}} = \begin{bmatrix} \partial h_{1k}^{(r)}/\partial f_1 & \partial h_{1k}^{(r)}/\partial \zeta_1 & \cdots & 0 & 0 \\ \vdots & \vdots & \ddots & \vdots & \vdots \\ 0 & 0 & \cdots & \partial h_{1k}^{(r)}/\partial f_m & \partial h_{1k}^{(r)}/\partial \zeta_m \end{bmatrix}^{\text{T}} \tag{B.8}$$

$$\frac{\partial \text{vec}_d(\overline{\boldsymbol{H}}_k^{(r)})}{\partial \boldsymbol{\theta}_m} = \overline{\left[ \frac{\partial \text{vec}_d(\boldsymbol{H}_k^{(r)})}{\partial \boldsymbol{\theta}_m} \right]} \tag{B.9}$$

The second-order derivative w.r.t. $\boldsymbol{\theta}_m$ can be calculated using the chain rule as

$$\frac{\partial^2 \mathcal{L}(\boldsymbol{\theta})}{\partial \boldsymbol{\theta}_m \partial \boldsymbol{\theta}_m^{\text{T}}} = \sum_{r=1}^{n_r} \sum_{k=1}^{N_f^{(r)}} \left[ \frac{\partial \text{vec}_d(\boldsymbol{H}_k^{(r)})}{\partial \boldsymbol{\theta}_m^{\text{T}}} \quad \frac{\partial \text{vec}_d(\overline{\boldsymbol{H}}_k^{(r)})}{\partial \boldsymbol{\theta}_m^{\text{T}}} \right] \cdot (\boldsymbol{I}_2 \otimes \boldsymbol{L}_d^{\text{T}}) \cdot \begin{bmatrix} \frac{\partial^2 \mathcal{L}(\boldsymbol{\theta})}{\partial \text{vec}(\boldsymbol{H}_k^{(r)}) \partial \text{vec}(\boldsymbol{H}_k^{(r)})^T} & \frac{\partial^2 \mathcal{L}(\boldsymbol{\theta})}{\partial \text{vec}(\overline{\boldsymbol{H}}_k^{(r)}) \partial \text{vec}(\boldsymbol{H}_k^{(r)})^T} \\ \frac{\partial^2 \mathcal{L}(\boldsymbol{\theta})}{\partial \text{vec}(\boldsymbol{H}_k^{(r)}) \partial \text{vec}(\overline{\boldsymbol{H}}_k^{(r)})^T} & \frac{\partial^2 \mathcal{L}(\boldsymbol{\theta})}{\partial \text{vec}(\overline{\boldsymbol{H}}_k^{(r)}) \partial \text{vec}(\overline{\boldsymbol{H}}_k^{(r)})^T} \end{bmatrix} \cdot$$

$$(\boldsymbol{I}_2 \otimes \boldsymbol{L}_d) \cdot \begin{bmatrix} \frac{\partial \text{vec}_d(\boldsymbol{H}_k^{(r)})}{\partial \boldsymbol{\theta}_m} \\ \frac{\partial \text{vec}_d(\overline{\boldsymbol{H}}_k^{(r)})}{\partial \boldsymbol{\theta}_m} \end{bmatrix} + \sum_{r=1}^{n_r} \sum_{k=1}^{N_f^{(r)}} \boldsymbol{G}_k^{(r)} \tag{B.10}$$

where

$$\boldsymbol{G}_k^{(r)} = \left\{ \left[ \frac{\partial \mathcal{L}(\boldsymbol{\theta})}{\partial \text{vec}(\boldsymbol{H}_k^{(r)})} \boldsymbol{L}_d \quad \frac{\partial \mathcal{L}(\boldsymbol{\theta})}{\partial \text{vec}(\overline{\boldsymbol{H}}_k^{(r)})} \boldsymbol{L}_d \right] \otimes \boldsymbol{I}_{2m} \right\} \frac{\partial}{\partial \boldsymbol{\theta}_m} \text{vec} \left( \frac{\partial \text{vec}_d(\boldsymbol{H}_k^{(r)})}{\partial \boldsymbol{\theta}_m^{\text{T}}} \quad \frac{\partial \text{vec}_d(\overline{\boldsymbol{H}}_k^{(r)})}{\partial \boldsymbol{\theta}_m^{\text{T}}} \right) \tag{B.11}$$

The term $\boldsymbol{G}_k^{(r)}$ looks complicated, but if it is expanded, one can find its $(p, q)$ term as

$$\boldsymbol{G}_k^{(r)}(p, q) = \sum_{ii} \left( \frac{\partial \mathcal{L}(\boldsymbol{\theta})}{\partial \boldsymbol{H}_{k,ii}^{(r)}} \cdot \frac{\partial^2 \boldsymbol{H}_{k,ii}^{(r)}}{\partial \boldsymbol{\theta}_p \partial \boldsymbol{\theta}_q} \quad \frac{\partial \mathcal{L}(\boldsymbol{\theta})}{\partial \overline{\boldsymbol{H}}_{k,ii}^{(r)}} \cdot \frac{\partial^2 \overline{\boldsymbol{H}}_{k,ii}^{(r)}}{\partial \boldsymbol{\theta}_p \partial \boldsymbol{\theta}_q} \right)$$

$$= \left[ \frac{\partial \mathcal{L}(\boldsymbol{\theta})}{\partial \text{vec}(\boldsymbol{H}_k^{(r)})} \quad \frac{\partial \mathcal{L}(\boldsymbol{\theta})}{\partial \text{vec}(\overline{\boldsymbol{H}}_k^{(r)})} \right] \cdot (\boldsymbol{I}_2 \otimes \boldsymbol{L}_d) \cdot \begin{bmatrix} \frac{\partial^2 \text{vec}_d(\boldsymbol{H}_k^{(r)})}{\partial \boldsymbol{\theta}_p \partial \boldsymbol{\theta}_q} \\ \frac{\partial^2 \text{vec}_d(\overline{\boldsymbol{H}}_k^{(r)})}{\partial \boldsymbol{\theta}_p \partial \boldsymbol{\theta}_q} \end{bmatrix} \tag{B.12}$$

with the expressions of $\frac{\partial^2 \text{vec}_d(\boldsymbol{H}_k^{(r)})}{\partial \boldsymbol{\theta}_p \partial \boldsymbol{\theta}_q} = \text{blkdiag}\{\nabla^2 h_{1k}^{(r)}, \dots, \nabla^2 h_{mk}^{(r)}\}$ ("bkldiag{·}" denotes a block diagonal matrix) and

$\nabla^2 h_{ik}^{(r)} = \begin{bmatrix} \partial^2 h_{ik}^{(r)}/\partial f_i \partial f_i & \partial^2 h_{ik}^{(r)}/\partial f_i \partial \zeta_i \\ \partial^2 h_{ik}^{(r)}/\partial \zeta_i \partial f_i & \partial^2 h_{ik}^{(r)}/\partial \zeta_i \partial \zeta_i \end{bmatrix}$ for $i = 1, 2, \dots, m$. For the complex conjugate term, one has

$$\frac{\partial^2 \text{vec}_d(\overline{\boldsymbol{H}}_k^{(r)})}{\partial \boldsymbol{\theta}_p \partial \boldsymbol{\theta}_q} = \overline{\left[ \frac{\partial^2 \text{vec}_d(\boldsymbol{H}_k^{(r)})}{\partial \boldsymbol{\theta}_p \partial \boldsymbol{\theta}_q} \right]} \tag{B.13}$$

For the first term of right-hand side of Eqn. (B.10), one has

$$\begin{cases}
\dfrac{\partial^2 \mathcal{L}(\boldsymbol{\theta})}{\partial \text{vec}(\boldsymbol{H}_k^{(r)}) \partial \text{vec}(\boldsymbol{H}_k^{(r)})^{\text{T}}} = \dfrac{\partial}{\partial \text{vec}(\boldsymbol{H}_k^{(r)})} S_e^{(r)-1} \text{vec}\left(-\boldsymbol{\Phi}^{\text{T}} \boldsymbol{S}^{(r)\text{T}} \overline{\boldsymbol{Y}_k^{(r)} \ddot{\boldsymbol{U}}_k^{(r)*}} \boldsymbol{\Lambda}^{(\mathcal{S}(r))} + \boldsymbol{\Phi}^{\text{T}} \boldsymbol{\Phi} \overline{\boldsymbol{H}}_k^{(r)} \boldsymbol{\Lambda}^{(\mathcal{S}(r))\text{T}} \overline{\ddot{\boldsymbol{U}}_k^{(r)} \ddot{\boldsymbol{U}}_k^{(r)*}} \boldsymbol{\Lambda}^{(\mathcal{S}(r))}\right) = \mathbf{0} \\
\dfrac{\partial^2 \mathcal{L}(\boldsymbol{\theta})}{\partial \text{vec}(\overline{\boldsymbol{H}}_k^{(r)}) \partial \text{vec}(\boldsymbol{H}_k^{(r)})^{\text{T}}} = S_e^{(r)-1} \left(\boldsymbol{\Lambda}^{(\mathcal{S}(r))\text{T}} \ddot{\boldsymbol{U}}_k^{(r)} \ddot{\boldsymbol{U}}_k^{(r)*} \boldsymbol{\Lambda}^{(\mathcal{S}(r))}\right) \otimes (\boldsymbol{\Phi}^{\text{T}} \boldsymbol{\Phi}) \\
\dfrac{\partial^2 \mathcal{L}(\boldsymbol{\theta})}{\partial \text{vec}(\boldsymbol{H}_k^{(r)}) \partial \text{vec}(\overline{\boldsymbol{H}}_k^{(r)})^{\text{T}}} = S_e^{(r)-1} \left(\boldsymbol{\Lambda}^{(\mathcal{S}(r))\text{T}} \overline{\ddot{\boldsymbol{U}}_k^{(r)} \ddot{\boldsymbol{U}}_k^{(r)*}} \boldsymbol{\Lambda}^{(\mathcal{S}(r))}\right) \otimes (\boldsymbol{\Phi}^{\text{T}} \boldsymbol{\Phi}) \\
\dfrac{\partial^2 \mathcal{L}(\boldsymbol{\theta})}{\partial \text{vec}(\overline{\boldsymbol{H}}_k^{(r)}) \partial \text{vec}(\overline{\boldsymbol{H}}_k^{(r)})^{\text{T}}} = \mathbf{0}
\end{cases} \quad (\text{B.14})$$

Substituting Eqn. (B.14) into Eqn. (B.10), Eqn. (B.10) can be simplified as

$$\dfrac{\partial^2 \mathcal{L}(\boldsymbol{\theta})}{\partial \boldsymbol{\theta}_m \partial \boldsymbol{\theta}_m^{\text{T}}} = \sum_{r=1}^{n_r} \sum_{k=1}^{N_f^{(r)}} 2 S_e^{(r)-1} \text{Re}\left\{\dfrac{\partial \text{vec}_d(\boldsymbol{H}_k^{(r)})}{\partial \boldsymbol{\theta}_m^{\text{T}}} \boldsymbol{L}_d^{\text{T}} [(\boldsymbol{\Lambda}^{(\mathcal{S}(r))\text{T}} \ddot{\boldsymbol{U}}_k^{(r)} \ddot{\boldsymbol{U}}_k^{(r)*} \boldsymbol{\Lambda}^{(\mathcal{S}(r))}) \otimes (\boldsymbol{\Phi}^{\text{T}} \boldsymbol{S}^{(r)\text{T}} \boldsymbol{S}^{(r)} \boldsymbol{\Phi})] \boldsymbol{L}_d \dfrac{\partial \text{vec}_d(\overline{\boldsymbol{H}}_k^{(r)})}{\partial \boldsymbol{\theta}_m}\right\} + \sum_{r=1}^{n_r} \sum_{k=1}^{N_f^{(r)}} \boldsymbol{G}_k^{(r)} \quad (\text{B.15})$$

2. Cross terms

We then derive the cross terms in the Hessian matrix $\nabla^2 \mathcal{L}(\boldsymbol{\theta})$, i.e., $\mathcal{L}^{(ab)} = \partial^2 \mathcal{L}(\boldsymbol{\theta})/\partial \boldsymbol{a} \partial \boldsymbol{b}$ for $\boldsymbol{a} \neq \boldsymbol{b}$.

(1) The second-order derivative of NLLF w.r.t. $S_e^{(r)}$ and $\boldsymbol{\Phi}$

$$\dfrac{\partial^2 \mathcal{L}(\boldsymbol{\theta})}{\partial S_e^{(r)} \partial \boldsymbol{\Phi}^{\text{T}}} = 2 S_e^{(r)-2} \text{Re}\left[\text{vec}\left(\sum_{k=1}^{N_f^{(r)}} \boldsymbol{S}^{(r)\text{T}} \widehat{\boldsymbol{Y}}_k^{(r)} \ddot{\boldsymbol{U}}_k^{(r)*} \boldsymbol{\Lambda}^{(\mathcal{S}(r))} \boldsymbol{H}_k^{(r)*}\right)\right] - 2 S_e^{(r)-2} \left[\text{Re}\left(\sum_{k=1}^{N_f^{(r)}} \boldsymbol{B}_k^{(r)}\right) \otimes (\boldsymbol{S}^{(r)\text{T}} \boldsymbol{S}^{(r)})\right] \text{vec}(\boldsymbol{\Phi}) \quad (\text{B.16})$$

(2) The second-order derivative of NLLF w.r.t. $S_e^{(r)}$ and $\boldsymbol{\Lambda}^{(s)}$

$$\dfrac{\partial^2 \mathcal{L}(\boldsymbol{\theta})}{\partial S_e^{(r)} \partial \boldsymbol{\Lambda}^{(s)\text{T}}} = 2 S_e^{(r)-2} \text{Re}\left[\text{vec}\left(\sum_{k=1}^{N_f^{(r)}} \ddot{\boldsymbol{U}}_k^{(r)} \widehat{\boldsymbol{Y}}_k^{(r)*} \boldsymbol{S}^{(r)} \boldsymbol{\Phi} \boldsymbol{H}_k^{(r)}\right)\right] - 2 S_e^{(r)-2} \left[\text{Re} \sum_{k=1}^{N_f^{(r)}} \boldsymbol{C}_k^{(r)\text{T}} \otimes (\ddot{\boldsymbol{U}}_k^{(r)} \ddot{\boldsymbol{U}}_k^{(r)*})\right] \text{vec}(\boldsymbol{\Lambda}^{(\mathcal{S}(r))}) \quad (\text{B.17})$$

(3) The second-order derivative of NLLF w.r.t. $S_e^{(r)}$ and $\boldsymbol{\theta}_m$

$$\dfrac{\partial^2 \mathcal{L}(\boldsymbol{\theta})}{\partial S_e^{(r)} \partial \boldsymbol{\theta}_m^{\text{T}}} = \sum_{k=1}^{N_f^{(r)}} \left[\dfrac{\partial \text{vec}_d(\boldsymbol{H}_k^{(r)})}{\partial \boldsymbol{\theta}_m^{\text{T}}} \quad \dfrac{\partial \text{vec}_d(\overline{\boldsymbol{H}}_k^{(r)})}{\partial \boldsymbol{\theta}_m^{\text{T}}}\right] \cdot (\boldsymbol{I}_2 \otimes \boldsymbol{L}_d^{\text{T}}) \cdot \begin{bmatrix} \dfrac{\partial^2 \mathcal{L}(\boldsymbol{\theta})}{\partial S_e^{(r)} \partial \text{vec}^{\text{T}}(\boldsymbol{H}_k^{(r)})} \\ \dfrac{\partial^2 \mathcal{L}(\boldsymbol{\theta})}{\partial S_e^{(r)} \partial \text{vec}^{\text{T}}(\overline{\boldsymbol{H}}_k^{(r)})} \end{bmatrix} \quad (\text{B.18})$$

where

$$\begin{cases}
\dfrac{\partial^2 \mathcal{L}(\boldsymbol{\theta})}{\partial S_e^{(r)} \partial \text{vec}^{\text{T}}(\boldsymbol{H}_k^{(r)})} = S_e^{(r)-2} \text{vec}\left[\boldsymbol{\Phi}^{\text{T}} \boldsymbol{S}^{(r)\text{T}} \left(\widehat{\boldsymbol{Y}}_k^{(r)} \ddot{\boldsymbol{U}}_k^{(r)*} - \boldsymbol{S}^{(r)} \boldsymbol{\Phi} \overline{\boldsymbol{H}}_k^{(r)} \boldsymbol{\Lambda}^{(\mathcal{S}(r))\text{T}} \overline{\ddot{\boldsymbol{U}}_k^{(r)} \ddot{\boldsymbol{U}}_k^{(r)*}}\right) \boldsymbol{\Lambda}^{(\mathcal{S}(r))}\right] \\
\dfrac{\partial^2 \mathcal{L}(\boldsymbol{\theta})}{\partial S_e^{(r)} \partial \text{vec}^{\text{T}}(\overline{\boldsymbol{H}}_k^{(r)})} = S_e^{(r)-2} \text{vec}\left[\boldsymbol{\Phi}^{\text{T}} \boldsymbol{S}^{(r)\text{T}} \left(\widehat{\boldsymbol{Y}}_k^{(r)} \ddot{\boldsymbol{U}}_k^{(r)*} - \boldsymbol{S}^{(r)} \boldsymbol{\Phi} \boldsymbol{H}_k^{(r)} \boldsymbol{\Lambda}^{(\mathcal{S}(r))\text{T}} \ddot{\boldsymbol{U}}_k^{(r)} \ddot{\boldsymbol{U}}_k^{(r)*}\right) \boldsymbol{\Lambda}^{(\mathcal{S}(r))}\right]
\end{cases} \quad (\text{B.19})$$

(4) The second-order derivative of NLLF w.r.t. $\boldsymbol{\Lambda}^{(s)}$ and $\boldsymbol{\Phi}$

$$\dfrac{\partial^2 L(\boldsymbol{\theta})}{\partial \boldsymbol{\Lambda}^{(s)} \partial \boldsymbol{\Phi}^{\text{T}}} = \sum_{r \in \mathcal{R}(s)} -2 S_e^{(r)-1} \sum_{k=1}^{N_f^{(r)}} \text{Re}(\overline{\boldsymbol{H}}_k^{(r)} \otimes \boldsymbol{S}^{(r)\text{T}} \widehat{\boldsymbol{Y}}_k^{(r)} \ddot{\boldsymbol{U}}_k^{(r)*}) +$$

$$\sum_{r \in \mathcal{R}(s)} 2 S_e^{(r)-1} \sum_k \text{Re}\left[\overline{\boldsymbol{H}}_k^{(r)} \otimes (\boldsymbol{S}^{(r)\text{T}} \boldsymbol{S}^{(r)} \boldsymbol{\Phi} \boldsymbol{H}_k^{(r)} \boldsymbol{\Lambda}^{(\mathcal{S}(r))\text{T}} \ddot{\boldsymbol{U}}_k^{(r)} \ddot{\boldsymbol{U}}_k^{(r)*}) + (\overline{\boldsymbol{H}}_k^{(r)} \boldsymbol{\Lambda}^{(\mathcal{S}(r))\text{T}} \overline{\ddot{\boldsymbol{U}}_k^{(r)} \ddot{\boldsymbol{U}}_k^{(r)*}}) \otimes (\boldsymbol{S}^{(r)\text{T}} \boldsymbol{S}^{(r)} \boldsymbol{\Phi} \boldsymbol{H}^{(r)}) \boldsymbol{K}_{ml}\right] \quad (\text{B.20})$$

where we have defined the commutation matrix $\boldsymbol{K}_{ml}$ consisting solely of zeros and ones such that $\text{vec}(\boldsymbol{\Lambda}^{(\mathcal{S}(r))\text{T}}) = \boldsymbol{K}_{ml} \text{vec}(\boldsymbol{\Lambda}^{(\mathcal{S}(r))})$.

(5) The second order derivative of NLLF w.r.t. $\boldsymbol{\Phi}$ and $\boldsymbol{\theta}_m$

$$\frac{\partial^2 \mathcal{L}(\boldsymbol{\theta})}{\partial \boldsymbol{\Phi} \partial \boldsymbol{\theta}_m^{\mathrm{T}}} = \sum_{r=1}^{n_r} \sum_{k=1}^{N_f^{(r)}} \left[ \frac{\partial \mathrm{vec}_d(\boldsymbol{H}_k^{(r)})}{\partial \boldsymbol{\theta}_m^{\mathrm{T}}} \quad \frac{\partial \mathrm{vec}_d(\overline{\boldsymbol{H}}_k^{(r)})}{\partial \boldsymbol{\theta}_m^{\mathrm{T}}} \right] \cdot (\boldsymbol{I}_2 \otimes \boldsymbol{L}_d^T) \begin{bmatrix} \frac{\partial^2 L(\boldsymbol{\theta})}{\partial \boldsymbol{\Phi} \partial \mathrm{vec}^{\mathrm{T}}(\boldsymbol{H}_k^{(r)})} \\ \frac{\partial^2 L(\boldsymbol{\theta})}{\partial \boldsymbol{\Phi} \partial \mathrm{vec}^{\mathrm{T}}(\overline{\boldsymbol{H}}_k^{(r)})} \end{bmatrix} \tag{B.21}$$

where,

$$\frac{\partial^2 \mathcal{L}(\boldsymbol{\theta})}{\partial \boldsymbol{\Phi} \partial \mathrm{vec}^{\mathrm{T}}(\boldsymbol{H}_k^{(r)})} = -S_e^{(r)^{-1}} \left( \boldsymbol{\Lambda}^{(\mathcal{S}(r))\mathrm{T}} \ddot{\boldsymbol{U}}_k^{(r)} \widehat{\boldsymbol{Y}}_k^{(r)*} \boldsymbol{S}^{(r)} \otimes \boldsymbol{I}_m \right) \boldsymbol{K}_{md} + S_e^{(r)^{-1}} \left( \boldsymbol{\Lambda}^{(\mathcal{S}(r))\mathrm{T}} \ddot{\boldsymbol{U}}_k^{(r)} \ddot{\boldsymbol{U}}_k^{(r)*} \boldsymbol{\Lambda}^{(\mathcal{S}(r))} \overline{\boldsymbol{H}}_k^{(r)} \right) \otimes \left( \boldsymbol{\Phi}^{\mathrm{T}} \boldsymbol{S}^{(r)\mathrm{T}} \right)$$
$$+ S_e^{(r)^{-1}} \left[ \left( \boldsymbol{\Lambda}^{(\mathcal{S}(r))\mathrm{T}} \ddot{\boldsymbol{U}}_k^{(r)} \ddot{\boldsymbol{U}}_k^{(r)*} \boldsymbol{\Lambda}^{(\mathcal{S}(r))} \overline{\boldsymbol{H}}_k^{(r)} \boldsymbol{\Phi}^{\mathrm{T}} \boldsymbol{S}^{(r)\mathrm{T}} \boldsymbol{S}^{(r)} \right) \otimes \boldsymbol{I}_m \right] \boldsymbol{K}_{md} \tag{B.22}$$

$$\frac{\partial^2 \mathcal{L}(\boldsymbol{\theta})}{\partial \boldsymbol{\Phi} \partial \mathrm{vec}^{\mathrm{T}}(\overline{\boldsymbol{H}}_k^{(r)})} = -S_e^{(r)^{-1}} \left( \boldsymbol{\Lambda}^{(\mathcal{S}(r))\mathrm{T}} \overline{\ddot{\boldsymbol{U}}_k^{(r)} \boldsymbol{Y}_k^*} \otimes \boldsymbol{I}_m \right) \boldsymbol{K}_{md} + S_e^{(r)^{-1}} \left( \boldsymbol{\Lambda}^{(\mathcal{S}(r))\mathrm{T}} \overline{\ddot{\boldsymbol{U}}_k^{(r)} \ddot{\boldsymbol{U}}_k^{(r)*}} \boldsymbol{\Lambda}^{(\mathcal{S}(r))} \boldsymbol{H}_k^{(r)} \right) \otimes \left( \boldsymbol{\Phi}^{\mathrm{T}} \boldsymbol{S}^{(r)\mathrm{T}} \right)$$
$$+ S_e^{(r)^{-1}} \left[ \left( \boldsymbol{\Lambda}^{(\mathcal{S}(r))\mathrm{T}} \overline{\ddot{\boldsymbol{U}}_k^{(r)} \ddot{\boldsymbol{U}}_k^{(r)*}} \boldsymbol{\Lambda}^{(\mathcal{S}(r))} \boldsymbol{H}_k^{(r)} \boldsymbol{\Phi}^{\mathrm{T}} \boldsymbol{S}^{(r)\mathrm{T}} \right) \otimes \boldsymbol{I}_m \right] \boldsymbol{K}_{md} \tag{B.23}$$

and the commutation matrix $\boldsymbol{K}_{md}$ is defined such that $\mathrm{vec}(\boldsymbol{\Phi}^{\mathrm{T}}) = \boldsymbol{K}_{md} \mathrm{vec}(\boldsymbol{\Phi})$.

(6) The second order derivative of NLLF w.r.t. $\boldsymbol{\Lambda}^{(s)}$ and $\boldsymbol{\theta}_m$

$$\frac{\partial^2 \mathcal{L}(\boldsymbol{\theta})}{\partial \boldsymbol{\Lambda}^{(s)} \partial \boldsymbol{\theta}_m^{\mathrm{T}}} = \sum_{r \in \mathcal{R}(s)} \sum_{k=1}^{N_f^{(r)}} \left[ \frac{\partial \mathrm{vec}_d(\boldsymbol{H}_k^{(r)})}{\partial \boldsymbol{\theta}_m^{\mathrm{T}}} \quad \frac{\partial \mathrm{vec}_d(\overline{\boldsymbol{H}}_k^{(r)})}{\partial \boldsymbol{\theta}_m^{\mathrm{T}}} \right] \cdot (\boldsymbol{I}_2 \otimes \boldsymbol{L}_d^T) \begin{bmatrix} \frac{\partial^2 L(\boldsymbol{\theta})}{\partial \boldsymbol{\Lambda}^{(s)} \partial \mathrm{vec}^{\mathrm{T}}(\boldsymbol{H}_k^{(r)})} \\ \frac{\partial^2 L(\boldsymbol{\theta})}{\partial \boldsymbol{\Lambda}^{(s)} \partial \mathrm{vec}^{\mathrm{T}}(\overline{\boldsymbol{H}}_k^{(r)})} \end{bmatrix} \tag{B.24}$$

where,

$$\frac{\partial^2 \mathcal{L}(\boldsymbol{\theta})}{\partial \boldsymbol{\Lambda}^{(s)} \partial \mathrm{vec}^{\mathrm{T}}(\boldsymbol{H}_k^{(r)})} = -S_e^{(r)^{-1}} \left( \boldsymbol{I}_m \otimes \boldsymbol{\Phi}^{\mathrm{T}} \boldsymbol{S}^{(r)\mathrm{T}} \overline{\boldsymbol{Y}_k^{(r)} \ddot{\boldsymbol{U}}_k^{(r)*}} \right) + S_e^{(r)^{-1}} \left[ \boldsymbol{I}_m \otimes \left( \boldsymbol{\Phi}^{\mathrm{T}} \boldsymbol{S}^{(r)\mathrm{T}} \boldsymbol{S}^{(r)} \boldsymbol{\Phi} \overline{\boldsymbol{H}}_k^{(r)} \boldsymbol{\Lambda}^{(\mathcal{S}(r))\mathrm{T}} \overline{\ddot{\boldsymbol{U}}_k^{(r)} \ddot{\boldsymbol{U}}_k^{(r)*}} \right) \right]$$
$$+ S_e^{(r)^{-1}} \left[ \left( \boldsymbol{\Lambda}^{(\mathcal{S}(r))\mathrm{T}} \ddot{\boldsymbol{U}}_k^{(r)} \ddot{\boldsymbol{U}}_k^{(r)*} \right) \otimes \left( \boldsymbol{\Phi}^{\mathrm{T}} \boldsymbol{S}^{(r)\mathrm{T}} \boldsymbol{S}^{(r)} \boldsymbol{\Phi} \overline{\boldsymbol{H}}_k^{(r)} \right) \right] \boldsymbol{K}_{ml} \tag{B.25}$$

$$\frac{\partial^2 \mathcal{L}(\boldsymbol{\theta})}{\partial \boldsymbol{\Lambda}^{(s)} \partial \mathrm{vec}^{\mathrm{T}}(\overline{\boldsymbol{H}}_k^{(r)})} = -S_e^{(r)^{-1}} \left( \boldsymbol{I}_m \otimes \boldsymbol{\Phi}^{\mathrm{T}} \boldsymbol{S}^{(r)\mathrm{T}} \boldsymbol{Y}_k^{(r)} \ddot{\boldsymbol{U}}_k^{(r)*} \right) + S_e^{(r)^{-1}} \left[ \boldsymbol{I}_m \otimes \left( \boldsymbol{\Phi}^{\mathrm{T}} \boldsymbol{S}^{(r)\mathrm{T}} \boldsymbol{S}^{(r)} \boldsymbol{\Phi} \boldsymbol{H}_k^{(r)} \boldsymbol{\Lambda}^{(\mathcal{S}(r))\mathrm{T}} \overline{\ddot{\boldsymbol{U}}_k^{(r)} \ddot{\boldsymbol{U}}_k^{(r)*}} \right) \right]$$
$$+ S_e^{(r)^{-1}} \left[ \left( \boldsymbol{\Lambda}^{(\mathcal{S}(r))\mathrm{T}} \overline{\ddot{\boldsymbol{U}}_k^{(r)} \ddot{\boldsymbol{U}}_k^{(r)*}} \right) \otimes \left( \boldsymbol{\Phi}^{\mathrm{T}} \boldsymbol{S}^{(r)\mathrm{T}} \boldsymbol{S}^{(r)} \boldsymbol{\Phi} \boldsymbol{H}_k^{(r)} \right) \right] \boldsymbol{K}_{ml} \tag{B.26}$$

# Appendix C  Derivatives of $h_k$

For the FRF $h_k = [(1 - \beta_k^2) - \mathbf{i}(2\zeta\beta_k)]^{-1}$ with $\beta_k = f/f_k$. The derivatives of $h_k$ w.r.t. $f$ and $\zeta$ are given by

$$h_k^{(f)} = 2h_k^2(\beta_k + \mathbf{i}\zeta)/f_k \tag{C.1}$$

$$h_k^{(\zeta)} = 2\mathbf{i} h_k^2 \beta_k / f_k \tag{C.2}$$

$$h_k^{(ff)} = 2h_k^3(3\beta_k^2 + 1 - 4\zeta^2 + 6\mathbf{i}\zeta\beta_k)/f_k^2 \tag{C.3}$$

$$h_k^{(\zeta\zeta)} = -8h_k^3 \beta_k^2 \tag{C.4}$$

$$h_k^{(f\zeta)} = 2\mathbf{i} h_k^3 (3\beta^2 + 1 + 2\mathbf{i}\zeta\beta)/f_k \tag{C.5}$$

$$h_k^{(\zeta f)} = 2\mathbf{i} h_k^3 (3\beta^2 + 1 + 2\mathbf{i}\zeta\beta)/f_k \tag{C.6}$$